\documentclass[
 	aip,
	11pt,
	a4paper,
	amsmath,
	amsfonts,
	amssymb,
	onecolumn,
	floatfix,
	superscriptaddress,
	nofootinbib
]{revtex4-2}

\usepackage{graphicx}
\usepackage[dvipsnames]{xcolor}
\usepackage{hyperref}
\usepackage{multirow}
\usepackage{hhline}
\usepackage{booktabs}
\usepackage{ulem}
\usepackage{dcolumn}
\usepackage{bm}
\usepackage[utf8]{inputenc}
\usepackage[T1]{fontenc}
\usepackage{mathptmx}
\usepackage{mathtools}
\usepackage{multirow}
\usepackage{hhline}
\usepackage{enumitem}
\usepackage{tcolorbox}
\usepackage{pdfpages}

\makeatletter
\AtBeginDocument{\let\LS@rot\@undefined}
\makeatother

\renewcommand{\emph}[1]{\textit{#1}}

\newcommand{\ex}[1]{\exp\left\lbrace #1\right\rbrace} % exponential function
\newcommand{\Int}[3]{\int\limits_{#1}^{#2}\text{d}#3\,} % integral
\newcommand{\Lim}[2]{{\lim\limits_{{#1} \rightarrow {#2}}}} % limit
\newcommand{\Sum}[2]{{\sum\limits_{#1}^{#2}}} % sum
\newcommand{\erw}[1]{\left\langle #1\right\rangle} % expectation/mean value
\newcommand{\pder}[2]{\frac{\partial #1}{\partial #2}} % partial derivative
\newcommand{\kb}{k_{\text{B}}} % Boltzmann constant

\begin{document}

% title
\title[How to define temperature in active systems?]{How to define temperature in active systems?}

% authors
\author{Lukas~Hecht}
\affiliation{ 
	Institute of Condensed Matter Physics, Department of Physics, Technical University of Darmstadt, Hochschulstraße 8, D-64289 Darmstadt, Germany
}

\author{Lorenzo~Caprini}
\affiliation{ 
	University of Rome La Sapienza, Piazzale Aldo Moro 5, 00185, Rome, Italy
}

\author{Hartmut~L\"owen}
\affiliation{ 
	Institut f\"ur Theoretische Physik II - Soft Matter, Heinrich-Heine-Universit\"at D\"usseldorf, Universit\"atsstraße 1, D-40225 D\"usseldorf, Germany
}

\author{Benno~Liebchen}
\email{benno.liebchen@pkm.tu-darmstadt.de}
\affiliation{ 
	Institute of Condensed Matter Physics, Department of Physics, Technical University of Darmstadt, Hochschulstraße 8, D-64289 Darmstadt, Germany
}

% date
\date{\today}

% abstract
\begin{abstract}
	We are used to measure temperature with a thermometer and we know from everyday life that different types of thermometers measure the same temperature. This experience can be based on equilibrium thermodynamics, which explains the equivalence of different possibilities to define temperature. In contrast, for systems out of equilibrium such as active matter, measurements performed with different thermometers can generally lead to different temperature values. In the present work, we systematically compare different possibilities to define temperature for active systems. Based on simulations and theory for inertial active Brownian particles, we find that different temperatures generally lead to different temperature values, as expected. Remarkably, however, we find that different temperatures not only lead to the same values near equilibrium (low P\'eclet number or high particle mass), but even far from equilibrium, several different temperatures approximately coincide. In particular, we find that the kinetic temperature, the configurational temperature, and temperatures based on higher moments of the velocity distribution constitute a class of temperatures that all assume very similar values over a wide parameter range. Notably, the effective temperature and temperatures exploiting the virial theorem, the Stokes-Einstein relation, or a harmonic confinement form a second class of temperatures whose values approximately coincide with each other but which strongly differ from those of the first class. Finally, we identify advantages and disadvantages of the different possibilities to define temperature and discuss their relevance for measuring the temperature of active systems.
\end{abstract}

\maketitle

\newpage
%%%%%%%%%%%%%%%%%%%%%%%%%%%%%%%%%%%%%%%%%%%%%%%%%%%%%%%%%%%%%%%%%%%%%%%%%%%%%%%%%%%%%%%%%%%%%%%%%%%%%%%%%%%%%
%%%%%%%%%%%%%%%%%%%%%%%%%%%%%%%%%%%%%%%%%%%%%%%%%%%%%%%%%%%%%%%%%%%%%%%%%%%%%%%%%%%%%%%%%%%%%%%%%%%%%%%%%%%%%
% INTRODUCTION
%%%%%%%%%%%%%%%%%%%%%%%%%%%%%%%%%%%%%%%%%%%%%%%%%%%%%%%%%%%%%%%%%%%%%%%%%%%%%%%%%%%%%%%%%%%%%%%%%%%%%%%%%%%%%
%%%%%%%%%%%%%%%%%%%%%%%%%%%%%%%%%%%%%%%%%%%%%%%%%%%%%%%%%%%%%%%%%%%%%%%%%%%%%%%%%%%%%%%%%%%%%%%%%%%%%%%%%%%%%
\section*{Introduction} \label{sec:intro}
``Temperature is a physical quantity that expresses quantitatively the attribute of hotness or coldness. Temperature is measured with a thermometer.''\cite{Wikipedia_Temperature} This is the temperature definition reported in the leading encyclopedia of our times.\cite{Wikipedia} Clearly, this notion of temperature is rather imprecise. Indeed, our sensation of hotness and coldness not only depends on temperature but also on the heat conductivity of the material we are touching. As an example, this can be experienced by touching a cold piece of wood and recognizing that it feels hotter than a piece of metal at the same temperature.\cite{Casas-Vazquez_RepProgPhys_2003}

There are many different ways to define temperature more precisely, and before we are taught thermodynamics, it may come as a surprise that in everyday life, different types of thermometers all essentially lead to the same result across a broad variety of environmental conditions. In particular, we may wonder why the reading of a liquid thermometer measuring the extension of a liquid agrees with the reading of an infrared thermometer that measures thermal radiation and even with that of a vapor-pressure thermometer that measures temperature through the vapor pressure of a liquid (exploiting the Clausius-Clapeyron equation). When learning statistical mechanics, we are in a position to understand that the universality of temperature and the link between different phenomena which are exploited by different types of thermometers exclusively hold true in thermodynamic equilibrium. In fact, the thermodynamic temperature can be linked to different observables in equilibrium systems. \cite{Jepps_PhysRevE_2000,Puglisi_PhysRep_2017} This leads to different equivalent possibilities to define temperature which exploit the equipartition theorem,\cite{Schroeder_Book_AnIntroductionToThermalPhysics_2021,Mello_AJPhys_2010} the virial theorem,\cite{Clausius_AnnPhys_1870} or fluctuation-dissipation relations for example.\cite{Berthier_JCP_2002,Berthier_PhysRevLett_2002,Geiss_ChemSystemsChem_2020,Sarracino_Chaos_2019} Alternatively, tracer particles can be used as a thermometer such that their properties can be linked to the thermodynamic temperature of the system in which the tracer particles are immersed.\cite{Baldovin_JStatMech_2017}

In principle, these and other definitions of temperature can all be generalized to non-equilibrium systems. In particular, classical irreversible thermodynamics grounds on the local-equilibrium hypothesis assuming that thermodynamic concepts like temperature may still be applied locally in non-equilibrium states.\cite{DeGroot_Book_NonEquilibriumThermodynamics_1984} However, when we are no longer near equilibrium and the local equilibrium hypothesis is invalid, the reading of a thermometer may (and typically will) depend on the details of the system under consideration (and may even be time-dependent).\cite{Mauro_JAmCeramSoc_2009,Petrelli_PhysRevE_2020} Such a situation is generally expected for systems where the relaxation times of certain degrees of freedom are long or if large persistent fluxes are present in the system, in particular, also for the large class of active matter systems containing self-propelled particles.\cite{Casas-Vazquez_RepProgPhys_2003,Puglisi_PhysRep_2017,Fodor_PhysRevLett_2016,Gaspard_Research_2020,Petrelli_PhysRevE_2020}

For such systems, we may wonder if it is sensible to define and speak of temperature at all. First, when touching a piece of glass or when putting our finger into a non-equilibrium liquid containing swimming bacteria, there is of course still a perception of hotness or coldness, and accordingly, it is tempting to introduce a measure to quantify our experience. Second, it may be instructive to explore when and by how much the different possibilities to define temperature, which we may use to quantify our experience, may deviate from each other. In particular, we may wonder if there are subsets in parameter space for which the reading of different thermometers would coincide. One might expect that different temperatures in active systems lead to strongly different temperature values for a system far from equilibrium, which can be quantified by measuring the total entropy production for example.\cite{Fodor_PhysRevLett_2016,Caprini_JStatMech_2019,Dabelow_PhysRevX_2019,Flenner_PhysRevE_2020}

In the present work, we comparatively explore different possibilities to define temperature for inertial active Brownian particles such as used in Refs.\ \onlinecite{Preisler_SoftMatter_2016,Omar_JCP_2023,Petrelli_PhysRevE_2020,Mandal_PhysRevLett_2019,Hecht_PRL_2022,Saw_PhysRevE_2023,Schwarzendahl_PRL_2022,Han_PNAS_2017,Cugliandolo_FluctNoiseLett_2019,Loi_PhysRevE_2008,Sanjay_PhysRevE_2022,Szamel_PhysRevE_2014,Levis_EPL_2015}. As expected, we find that different temperatures lead to results that depend on the details of the considered non-equilibrium system, and in general, that all obtained temperature values deviate from each other. However, perhaps surprisingly, we identify parameter regimes where different temperatures provide consistent results even far from equilibrium. This applies in particular to regimes in which the active particles are heavy or in which their rotational diffusion is fast, and it has been previously found that within these regimes, an active system behaves as an effective equilibrium system.\cite{Caprini_JCP_2023,Caprini_JCP_2021,Omar_JCP_2023} Indeed, we find that within this regime, also the considered temperatures lead to similar temperature values independently of the values of all other dimensionless parameters that control the dynamics of the active particles. Interestingly, we find that the different temperatures which we have compared can be sorted in two classes: The first one shows a strong mass dependence (and scales linearly with the mass in a wide parameter regime) and the second one is almost mass independent. We show that these two classes can approximately be matched by rescaling with the particle mass. This finding has important consequences for the calculation of temperature in active systems, as we shall see.

The article is organized as follows: First, we introduce the active particle model and the different possibilities to define temperature, and we summarize known analytical results. Second, we present new numerical results based on Brownian dynamics simulations of inertial active Brownian particles. Third, we discuss the advantages and disadvantages of the presented temperatures, and finally, we conclude our work.

%%%%%%%%%%%%%%%%%%%%%%%%%%%%%%%%%%%%%%%%%%%%%%%%%%%%%%%%%%%%%%%%%%%%%%%%%%%%%%%%%%%%%%%%%%%%%%%%%%%%%%%%%%%%%
%%%%%%%%%%%%%%%%%%%%%%%%%%%%%%%%%%%%%%%%%%%%%%%%%%%%%%%%%%%%%%%%%%%%%%%%%%%%%%%%%%%%%%%%%%%%%%%%%%%%%%%%%%%%%
% RESULTS
%%%%%%%%%%%%%%%%%%%%%%%%%%%%%%%%%%%%%%%%%%%%%%%%%%%%%%%%%%%%%%%%%%%%%%%%%%%%%%%%%%%%%%%%%%%%%%%%%%%%%%%%%%%%%
%%%%%%%%%%%%%%%%%%%%%%%%%%%%%%%%%%%%%%%%%%%%%%%%%%%%%%%%%%%%%%%%%%%%%%%%%%%%%%%%%%%%%%%%%%%%%%%%%%%%%%%%%%%%%

%%%%%%%%%%%%%%%%%%%%%%%%%%%%%%%%%%%%%%%%%%%%%%%%%%%%%%%%%%%%%%%%%%%%%%%%%%%%%%%%%%%%%%%%%%%%%%%%%%%%%%%%%%%%%
% MODELS
%%%%%%%%%%%%%%%%%%%%%%%%%%%%%%%%%%%%%%%%%%%%%%%%%%%%%%%%%%%%%%%%%%%%%%%%%%%%%%%%%%%%%%%%%%%%%%%%%%%%%%%%%%%%%
\section*{Model}
In this work, we consider inertial active particles modeled by the active Brownian particle (ABP) model \cite{Hecht_ArXiv_2021,Romanczuk_EurPhysJSpecialTopics_2012,Loewen_JCP_2020,Digregorio_PhysRevLett_2018,Shaebani_NatRevPhys_2020} in two spatial dimensions. The ABP model is a ``dry'' model, i.e., self-propulsion is modeled effectively and the solvent solely acts as a thermal bath that leads to fluctuations in the equations of motion.\cite{Hecht_ArXiv_2021} Within the ABP model, an active particle is represented by a (slightly soft) sphere of diameter $\sigma$ with mass $m$ and moment of inertia $I=m\sigma^2/10$ (corresponding to a rigid sphere). The particles feature an effective self-propulsion force $\mathbf{F}_{\text{SP},i}=\gamma_{\text{t}}v_{0}\mathbf{p}_i(t)$, where $v_{0},\mathbf p_i$ denote the (terminal) self-propulsion speed and the self-propulsion direction $\mathbf{p}_i$ of the $i$-th active particle ($i=1,2,...,N$), respectively. Its position $\mathbf{r}_i$ evolves as $\text{d}\mathbf{r}_i/\text{d}t=\mathbf{v}_i$, and its velocity $\mathbf{v}_i$ evolves according to
\begin{equation}
	m\frac{\text{d}\mathbf{v}_i}{\text{d} t} =-\gamma_{\text{t}}\mathbf{v}_i+\gamma_{\text{t}}v_{0}\mathbf{p}_i - \sum_{\substack{j=1\\j\neq i}}^{N}\boldsymbol{\nabla}_{\mathbf{r}_i}u\left(r_{ij}\right) + \mathbf{F}_{\text{ext},i} + \sqrt{2 k_{\rm B} T_{\text{b}}\gamma_{\text{t}}}\boldsymbol{\xi}_i.\label{eq:active-trans-leq}
\end{equation}
Here, $T_{\text{b}}$ represents the bath temperature, $\boldsymbol{\xi}_{i}$ denotes Gaussian white noise with zero mean and unit variance, and $\gamma_{\text{t}}$ denotes the translational drag coefficient. The particles may interact through a two-body interaction potential $u(r_{ij})$, $r_{ij}=\left\lvert\mathbf{r}_i-\mathbf{r}_j\right\rvert$ and may be subject to an additional external force $\mathbf{F}_{\text{ext},i}=-\boldsymbol{\nabla}_{\mathbf{r}_i}U_{\text{ext}}(\mathbf{r}_i)$ originating from an external potential $U_{\text{ext}}$. The self-propulsion direction $\mathbf{p}_i(t)$ can be expressed in terms of the orientation angle $\phi_i(t)$ as $\mathbf{p}_i(t)=(\cos\phi_i(t),\sin\phi_i(t))$.\cite{Romanczuk_EurPhysJSpecialTopics_2012,Mandal_PhysRevLett_2019,Loewen_JCP_2020,Gutierrez-Martinez_JCP_2020,Sandoval_PhysRevE_2020,Su_NewJPhys_2021,Hecht_PRL_2022} It evolves in time according to $\text{d}\phi_i/\text{d}t = \omega_i$, where the angular velocity $\omega_i$ evolves as
\begin{equation}
	I\frac{\text{d}\omega_i}{\text{d} t} = -\gamma_{\text{r}}\omega_i+\sqrt{2 k_{\rm B} T_{\text{b}}\gamma_{\text{r}}}\eta_{i}.\label{eq:abp-rot-leq}
\end{equation}
Here, $\gamma_{\text{r}}$ denotes the rotational drag coefficient and $\eta_{i}$ denotes Gaussian white noise with zero mean and unit variance. In the overdamped limit, it yields
\begin{equation}
	\frac{\text{d}\phi_i}{\text{d}t}=\sqrt{\frac{2}{\tau_{\text{p}}}}\eta_{i},\label{eq:abp-rot-od}
\end{equation}
where $\tau_{\text{p}}=1/D_{\text{r}}$ denotes the persistence time and $D_{\text{r}}=k_{\text{B}}T_{\text{b}}/\gamma_{\text{r}}$ the rotational diffusion coefficient. Here, $k_{\text{B}}$ is the Boltzmann constant.

%%%%%%%%%%%%%%%%%%%%%%%%%%%%%%%%%%%%%%%%%%%%%%%%%%%%%%%%%%%%%%%%%%%%%%%%%%%%%%%%%%%%%%%%%%%%%%%%%%%%%%%%%%%%%
% TEMPERATURE DEFINITIONS
%%%%%%%%%%%%%%%%%%%%%%%%%%%%%%%%%%%%%%%%%%%%%%%%%%%%%%%%%%%%%%%%%%%%%%%%%%%%%%%%%%%%%%%%%%%%%%%%%%%%%%%%%%%%%
\section*{Possibilities to define temperature}
Let us first discuss the different possibilities under consideration to define temperature. In general, we can distinguish three different approaches to define temperature: First, one can define temperature based on the fluctuations of the particle velocity, which is a very common approach in the field of granular particles.\cite{Goldhirsch_PowTech_2008,Puglisi_PhysRevE_2002,Kumaran_PhysRevE_1998,Baldassarri_JPhysCondensMatter_2005} Second, it is possible to define temperature based on fluctuations in particle positions.\cite{Szamel_PhysRevE_2014} The third approach takes inspiration from glassy systems and exploits fluctuation-dissipation relations.\cite{Cugliandolo_PhysRevE_1997,Leonardo_PhysRevLett_2000,Kob_EurPhysJB_2000} In the following, we briefly introduce the different possibilities to define temperature as used in this work and summarize some known analytical results. The considered temperatures are summarized in Tab.\ \ref{tab:temperature-definitions}.

Before introducing the different possibilities to define temperature, let us recap some general concepts known from equilibrium physics. In particular, let us consider an equilibrium system of $N$ particles in three spatial dimensions and let $\boldsymbol{\Gamma}=(\Gamma_1,...,\Gamma_{6N})=(p_1,...,p_{3N}, q_1,...,q_{3N})$ be the phase-space vector representing the spatial coordinates $q_i$ and the conjugate momenta $p_i$. Furthermore, let the system be described by the Hamiltonian $\mathcal{H}(\boldsymbol{\Gamma} )=\sum_i p_i^2/(2m) + V(\lbrace q_j\rbrace)$, where $m$ denotes the mass of the particles and $V$ the potential energy of the system. Based on the standard thermodynamic relation $1/T=\text{d}S(E)/\text{d}E$ with entropy $S(E)$ and energy $E$, one can show that the thermodynamic temperature can be calculated as \cite{Jepps_PhysRevE_2000}
\begin{equation}
	k_{\text{B}}T=\frac{\langle\boldsymbol{\nabla}\mathcal{H}\cdot\mathbf{B}(\boldsymbol{\Gamma})\rangle}{\langle\boldsymbol{\nabla}\cdot\mathbf{B}(\boldsymbol{\Gamma})\rangle},\label{eq:general-temp}
\end{equation}
where $\mathbf{B}(\boldsymbol{\Gamma})$ is an arbitrary vector field with $0<|\langle\boldsymbol{\nabla}\mathcal{H}\cdot\mathbf{B}(\boldsymbol{\Gamma})\rangle|,|\langle\boldsymbol{\nabla}\cdot\mathbf{B}(\boldsymbol{\Gamma})\rangle|<\infty$ and $\boldsymbol{\nabla}$ is the gradient operator in the $3N$-dimensional space. Furthermore, $\langle\boldsymbol{\nabla}\mathcal{H}\cdot\mathbf{B}(\boldsymbol{\Gamma})\rangle$ must grow slower than $e^N$ in the thermodynamic limit.\cite{Jepps_PhysRevE_2000} Note that for $\mathbf{B}(\boldsymbol{\Gamma})=(0,...,\Gamma_i,...,0)$, we obtain the generalized equipartition theorem $k_{\text{B}}T=\langle\Gamma_i\partial\mathcal{H}/\partial\Gamma_i\rangle$. If $\Gamma_i=p_i$, we recover the equipartition theorem $k_{\text{B}}T=\langle p_i^2/m\rangle$,\cite{Mello_AJPhys_2010} which we will exploit for some temperature definitions below. In turn, if $\Gamma_i$ is a coordinate $q_i$, we obtain the Clausius virial theorem $k_{\text{B}}T=-\langle q_iF_i\rangle$.\cite{Clausius_AnnPhys_1870} From the general expression in Eq.\ (\ref{eq:general-temp}), we can directly derive different temperatures such as the kinetic temperature and the configurational temperature as shown below.

\renewcommand{\arraystretch}{1.4}
\begin{table}
	\centering
	\caption{\textbf{Temperature definitions.} Summary of different possibilities to define temperature for a system of (inertial) ABPs in $d$ spatial dimensions.}
	\begin{tabular}{l|l|p{7cm}|l|l}
		\textbf{Symbol} & \textbf{Name} & \textbf{Definition} & \textbf{References} & \textbf{Comments}\\
		\hhline{=====}
		\multicolumn{5}{l}{\textit{Velocity-based definitions}} \\
		\hhline{=====}
		$T_{\rm kin}$ & kinetic & $\frac{1}{2Nd}\sum_{i=1}^{N}m\left\langle\left(\mathbf{v}_i-\erw{\mathbf{v}}\right)^2\right\rangle$ & \cite{DeKarmakar_PhysRevE_2020,Caprini_JCP_2020,Hecht_PRL_2022,Mandal_PhysRevLett_2019,Marconi_SciRep_2017,Petrelli_PhysRevE_2020,Schiltz-Rouse_PhysRevE_2023,Marconi_NewJPhys_2021,Maggi_SoftMatter_2021} & Eq.\ (\ref{eq:kinetic-temperature})\\
		\hline
		$T_{\rm kin4}$ & fourth-moment kinetic & $\frac{1}{2}m\sqrt{\frac{4}{Nd(d+2)}\sum_{i=1}^{N}\left\langle\left(\mathbf{v}_i-\erw{\mathbf{v}}\right)^4\right\rangle}$ &  & Eq.\ (\ref{eq:fourth-kinetic-temperature})\\
		\hline
		$T_{\rm MB}$ & Maxwell-Boltzmann & $\sqrt{\frac{m_{\text{tracer}}}{2\pi\kb T_{\text{MB}}}}\ex{-\frac{m_{\text{tracer}}v_i^2}{2\kb T_{\text{MB}}}},~i=x,y,z$ & \cite{Loi_PhysRevE_2008} & Eq.\ (\ref{eq:Maxwell-Boltzmann-temperature})\\
		\hline
		\multicolumn{5}{l}{\textit{Position-based definitions}} \\
		\hhline{=====}
		$T_{\rm vir}$ & virial & $\gamma_{\text{t}}\lim\limits_{t\rightarrow\infty}\partial_t\text{MSD}(t)+$ $\frac{1}{2Nd}\sum_{i=1}^{N} \left\langle\mathbf{r}_i\cdot\mathbf{F}_{\text{ext},i} - \sum_{j<i}\mathbf{r}_{ij}\cdot\mathbf{F}_{ij} - \gamma_{\text{t}}v_0\mathbf{r}_i\cdot\mathbf{p}_i \right\rangle$ & & Eq.\ (\ref{eq:virial-temp-abps})\\
		\hline
		$T_{\rm osc}$ & oscillator & $k \erw{x^2}$ & \cite{Hecht_PRL_2022} & Eq.\ (\ref{eq:oscillator-temperature})\\
		\hline
		$T_{\rm conf}$ & configurational & $\frac{\erw{\left(\boldsymbol{\nabla} U_{\text{tot}}\right)^2}}{\erw{\nabla^2 U_{\text{tot}}}}$ & \cite{Saw_PhysRevE_2023,Saw_PhysRevE_2023_2} & Eq.\ (\ref{eq:Tconf}) \\
		\hline
		\multicolumn{5}{l}{\textit{Dynamics-based definitions}} \\
		\hhline{=====}
		$T_{\rm Ein}$ & Einstein & $\gamma_{\text{t}} D_{\text{eff}}$ & & Eq.\ (\ref{eq:einstein-temperature})\\
		\hline
		$T_{\rm eff}$ & effective & $\lim\limits_{t\gg 1}\frac{{\rm MSD}(t)}{2d\chi(t)}$ & \cite{Cugliandolo_FluctNoiseLett_2019,Petrelli_PhysRevE_2020,Loi_SoftMatter_2011} & Eq.\ (\ref{eq:Teff})\\
		\bottomrule
	\end{tabular}
	\label{tab:temperature-definitions}
\end{table}
\renewcommand{\arraystretch}{1.0}

\newpage
\subsection*{Velocity-based definitions}
Velocity fluctuations can be used to define temperature either based on the velocities of the active particles themselves or based on the velocity distribution of tracer particles that are suspended in a bath of active particles. Here, we consider the following possibilities to define temperature based on velocities:
\begin{enumerate}
	\item[(1)] \textbf{Kinetic temperature:} Starting from Eq.\ (\ref{eq:general-temp}), we can derive the kinetic temperature by choosing $\mathbf{B}(\boldsymbol{\Gamma})=(0,...,0,p_1,..., p_{3N})$, which yields \cite{Himpel_PhysRevE_2019,Han_JCP_2005} 
	\begin{equation}
		k_{\text{B}}T=\left\langle\frac{1}{3N}\sum_{i=1}^{3N}\frac{p_i^2}{m}\right\rangle,\label{eq:general-Tkin}
	\end{equation}
	where $p_i=m v_i$ is the momentum of the $i$-th particle. When we consider the average translational kinetic energy $E_{\rm kin}=\frac{1}{2}\sum_{i=1}^{N}m_i\left(\mathbf{v}_i-\erw{\mathbf{v}}\right)\cdot\left(\mathbf{v}_i-\erw{\mathbf{v}}\right)$ (subtracting any possible drift velocity $\erw{\mathbf{v}}$), we can write the kinetic temperature as
	\begin{equation}
		\kb T_{\rm kin}\coloneqq\frac{1}{Nd}\sum_{i=1}^{N}m\left\langle\left(\mathbf{v}_i-\erw{\mathbf{v}}\right)\cdot\left(\mathbf{v}_i-\erw{\mathbf{v}}\right)\right\rangle, \label{eq:kinetic-temperature}
	\end{equation}	
	where $d$ is the spatial dimension. The kinetic temperature is commonly used in the field of granular particles \cite{Grasselli_EPJE_2015,Komatsu_PhysRevX_2015,Prevost_PhysRevE_2004,Lobkovsky_EurPhysJSpecialTopics_2009,Melby_JPhysCondensMatter_2005,VegaReyes_PhysRevE_2008,Roeller_PhysRevLett_2011,Feitosa_PhysRevLett_2002, Campbell_PowderTechnology_2006} and complex plasmas,\cite{Ivelv_Book_ComplexPlasmasAndColloidalDispersions_2012,Ivlev_PhysRevX_2015} and it is equal to the thermodynamic temperature in equilibrium systems where $\langle\mathbf{v}\rangle = 0$ and Eq.\ (\ref{eq:kinetic-temperature}) coincides with Eq.\ (\ref{eq:general-Tkin}).\cite{Schroeder_Book_AnIntroductionToThermalPhysics_2021} It has also been frequently used as a well-defined temperature definition in active systems.\cite{DeKarmakar_PhysRevE_2020,Caprini_JCP_2020,Hecht_PRL_2022,Mandal_PhysRevLett_2019,Marconi_SciRep_2017,Petrelli_PhysRevE_2020,Schiltz-Rouse_PhysRevE_2023,Marconi_NewJPhys_2021,Maggi_SoftMatter_2021} Note that the kinetic temperature is proportional to the mass of the particles. This has an important implication for the kinetic temperature of (inertial) active particles: Free non-interacting active particles move with their terminal speed $v_0$ in the steady state independently of their mass. Therefore, their kinetic temperature strongly depends on their mass as demonstrated in Fig.\ \ref{fig:tkin-sketch}.
	
	For the kinetic temperature in active systems, some analytical results are known. For example, the kinetic temperature of free non-interacting ABPs [$u=0$ and $U_{\text{ext}}=0$ in Eq.\ (\ref{eq:active-trans-leq})] and similarly of active Ornstein-Uhlenbeck particles (AOUPs; see Appendix A) can be written as \cite{Caprini_JCP_2021}
	\begin{equation}
		\kb T_{\text{kin}}= \kb T_{\text{b}} + m v_0^2\alpha, \label{eq:Tkin-analytic-free}
	\end{equation} 
	where $m$ is the mass of the particles and the dimensionless coefficient $\alpha$ is given by
	\begin{equation}
		\alpha = \frac{\tau_{\text{p}}\gamma_{\text{t}}/m}{1 + \tau_{\text{p}}\gamma_{\text{t}}/m}. \label{eq:Tkin-alpha-free}
	\end{equation}
	The first term in Eq.\ (\ref{eq:Tkin-analytic-free}) is the bath temperature that determines the strength of the Brownian noise. The second term has a pure non-equilibrium origin and disappears in equilibrium. Note that Eq.\ (\ref{eq:Tkin-analytic-free}) can be obtained based on the AOUP model leading to similar results as for ABPs (see Appendix A). Further analytical results are shown in Appendix B.
	
	The previous definition of kinetic temperature is based on the second moment of the velocity distribution. Similarly, one can use higher moments to define a variant of the kinetic temperature. Exemplarily, we introduce a temperature based on the fourth moment. In particular, we obtain
	\begin{equation}
		\kb T_{\rm kin4}\coloneqq\frac{1}{2}m\sqrt{\frac{4}{Nd(d+2)}\sum_{i=1}^{N}\left\langle\left(\mathbf{v}_i-\erw{\mathbf{v}}\right)^4\right\rangle}, \label{eq:fourth-kinetic-temperature}
	\end{equation}
	where $d=1,2,3$ is the spatial dimension. This temperature is again equal to the bath temperature in equilibrium, i.e., for $v_0=0$ in Eq.\ (\ref{eq:active-trans-leq}). Note that $T_{\rm kin}=T_{\rm kin4}$ if the velocity distribution $\mathcal{P}(v_i),~i=x,y,z$ is Gaussian, i.e., if $\mathcal{P}(|\mathbf{v}|)$ follows the Maxwell-Boltzmann distribution.
	
	\begin{figure}
		\centering
		\includegraphics[width=0.5\linewidth]{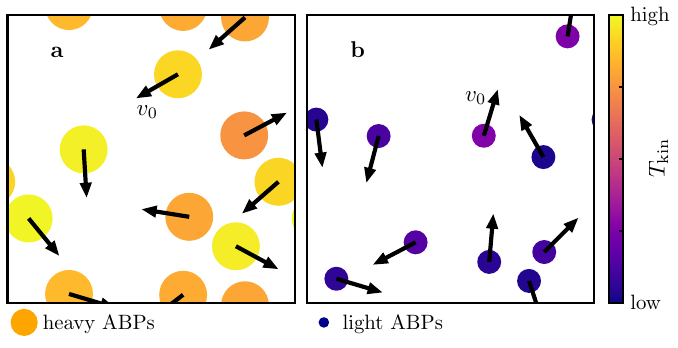}
		\caption{\textbf{Kinetic temperature.} Schematic visualization of the kinetic temperature of active particles moving with a terminal self-propulsion speed $v_0$. \textbf{a} Heavy active particles feature a large kinetic temperature and \textbf{b} light active particles have a low kinetic temperature. The color denotes the kinetic temperature, the black arrows the velocity of the particles.}
		\label{fig:tkin-sketch}
	\end{figure}

	\item[(2)] \textbf{Maxwell-Boltzmann temperature:} For free non-interacting particles of mass $m$ in a classical equilibrium gas at temperature $T$, the Maxwell-Boltzmann distribution reads \cite{Maxwell_LEDPhilMagJSci_1860,Boltzmann_WienBer_1877}
	\begin{equation}
		\mathcal{P}(v_x,v_y,v_z)=\left(\frac{m}{2\pi\kb T}\right)^{3/2}\ex{-\frac{m(v_x^2+v_y^2+v_z^2)}{2\kb T}}, \label{eq:Maxwell-Boltzmann-Distribution}
	\end{equation}
	i.e., each velocity component $v_i,~i=x,y,z$ is Gaussian distributed:
	\begin{equation}
		\mathcal{P}(v_i)=\sqrt{\frac{m}{2\pi\kb T}}\ex{-\frac{mv_i^2}{2\kb T}}.
	\end{equation}
	Within an equilibrium system, the Maxwell-Boltzmann distribution can be exploited to determine the temperature of the system by measuring the velocity distribution of the particles. However, since the velocity of active particles is generally not Maxwell-Boltzmann distributed, this procedure is not directly applicable to active systems. In turn, one could use passive tracer particles as a thermometer (Fig.\ \ref{fig:tmb-sketch}). While passive particles immersed in an active bath can be out of equilibrium, there are some parameter regimes, in which their velocity distribution approximately has a Maxwell-Boltzmann shape.\cite{Shea_SoftMatter_2022,Hecht_NatComm_2024} Therefore, their velocity distribution provides an approximate measure for the temperature of the active particles, the Maxwell-Boltzmann temperature $T_{\text{MB}}$ (Fig.\ \ref{fig:tmb-sketch}b) defined via
	\begin{equation}
		\mathcal{P}(v_i)=\sqrt{\frac{m_{\text{tracer}}}{2\pi\kb T_{\text{MB}}}}\ex{-\frac{m_{\text{tracer}}v_i^2}{2\kb T_{\text{MB}}}},~i=x,y,z,\label{eq:Maxwell-Boltzmann-temperature}
	\end{equation}
	where $m_{\text{tracer}}$ denotes the mass of the tracer particle. Note that the Maxwell-Boltzmann temperature depends on the tracer mass and on the interactions between the active particles and the tracer. In fact, the tracer should follow the slow dynamics of the active system, which is only guaranteed if the tracer is sufficiently heavy.\cite{Loi_PhysRevE_2008} In addition, the tracer has to be small enough to not affect the structure of the active system. Note that there are parameter regimes in which the velocity distribution is not Gaussian anymore,\cite{Hecht_NatComm_2024} and therefore, the Maxwell-Boltzmann temperature cannot be calculated (see inset in Fig.\ \ref{fig:tmb-sketch}b). Hence, we do not show the Maxwell-Boltzmann temperature in the numerical results below.
	
	\begin{figure}
		\centering
		\includegraphics[width=0.5\linewidth]{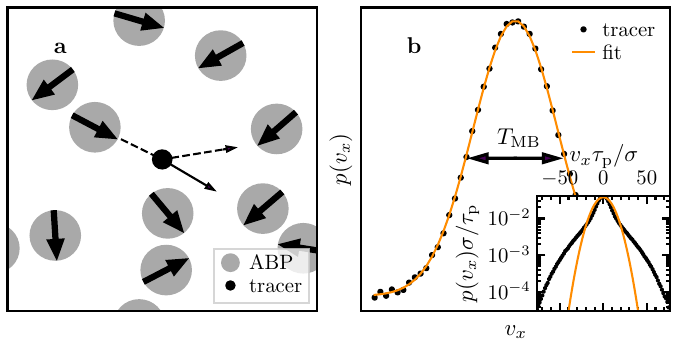}
		\caption{\textbf{Maxwell-Boltzmann temperature.} \textbf{a} A passive tracer particle (black) that interacts with the surrounding active particles (gray) is used as a thermometer for the active particles. \textbf{b} Velocity distribution (black dots) of passive tracer particles in a bath of active particles. The Maxwell-Boltzmann temperature $T_{\text{MB}}$ is obtained from the variance of a Gaussian (yellow line) fitted to the tracer velocity distribution. The inset shows the velocity distribution for the passive particles in a mixture of overdamped ABPs ($m/(\gamma_{\text{t}}\tau_{\text{p}})=5\times 10^{-5}$) and inertial passive Brownian particles ($m/(\gamma_{\text{t}}\tau_{\text{p}})=5\times 10^{-2}$) at $\text{Pe}=100$, $\varphi_{\text{tot}}=0.5$, and $x_{\text{a}}=0.9$, where $x_{\text{a}}$ denotes the fraction of active particles. In this parameter regime, the velocity distribution is clearly non-Gaussian, and therefore, the Maxwell-Boltzmann temperature cannot be sensibly calculated. The data has been taken from Ref.\ \onlinecite{Hecht_NatComm_2024}.}
		\label{fig:tmb-sketch}
	\end{figure}
	
\end{enumerate}

\subsection*{Position-based definitions}
We now introduce different possibilities to define temperature based on the positions of active particles and tracers.

\begin{enumerate}[label=\textbf{(\alph*)}]
	\item[(3)] \textbf{Virial temperature:} The virial theorem connects the average kinetic energy of a system to its average potential energy by $\frac{1}{2}\sum_{i=1}^{N}m\left\langle\mathbf{v}_i\cdot\mathbf{v}_i\right\rangle = -\frac{1}{2}\sum_{i=1}^{N}\left\langle\mathbf{r}_i\cdot\mathbf{F}_i\right\rangle$	and was first introduced by R.\ Clausius in 1870.\cite{Clausius_AnnPhys_1870} Here, $\mathbf{F}_i$ denotes the total force acting on the $i$-th particle. In equilibrium, the average virial $\langle\mathcal{V}\rangle=-\frac{1}{2}\sum_{i=1}^{N}\left\langle\mathbf{r}_i\cdot\mathbf{F}_i\right\rangle$ can be connected to the thermodynamic temperature $T$ of the system by applying the equipartition theorem as already done for the definition of the kinetic temperature. This leads to the virial temperature defined as
	\begin{equation}
		\kb T_{\rm vir}\coloneqq\frac{2}{Nd}\langle\mathcal{V}\rangle,\label{eq:virial-temperature}
	\end{equation}
	where $d$ denotes the spatial dimension of the system and $N$ is again the number of active particles.\cite{Becker_Book_TheoryOfHeat_1967} For inertial ABPs [Eqs.\ (\ref{eq:active-trans-leq}) and (\ref{eq:abp-rot-leq})] in the steady state, the virial temperature can be written as (see Appendix C for details)
	\begin{equation}
		\kb T_{\rm vir}^{\text{ABP}}=\gamma_{\text{t}}\lim\limits_{t\rightarrow\infty}\partial_t\text{MSD}(t) + \frac{1}{2Nd}\sum_{i=1}^{N} \left\langle\mathbf{r}_i\cdot\mathbf{F}_{\text{ext},i} - \sum_{j<i}\mathbf{r}_{ij}\cdot\mathbf{F}_{ij} - \gamma_{\text{t}}v_0\mathbf{r}_i\cdot\mathbf{p}_i \right\rangle\label{eq:virial-temp-abps}
	\end{equation}
	with the mean-square displacement $\text{MSD}(t)=\langle\mathbf{r}_i(t)^2\rangle$ (assuming $\mathbf{r}_i(0)=0$), interaction force $\mathbf{F}_{ij}$, and spatial dimension $d$. Hence, the virial temperature does only require information about the positions and the forces but not about the velocities of the particles. Therefore, it is also applicable to simulations in the overdamped limit. Note that for free ABPs, the first term is equal to $\gamma_{\text{t}}D_{\text{eff}}$, where $D_{\text{eff}}$ is their effective diffusion coefficient, which we obtain from their long-time MSD. All other contributions are directly calculated from the particle trajectories by averaging over time in the steady state.
	
	\begin{figure}
		\centering
		\includegraphics[width=0.6\linewidth]{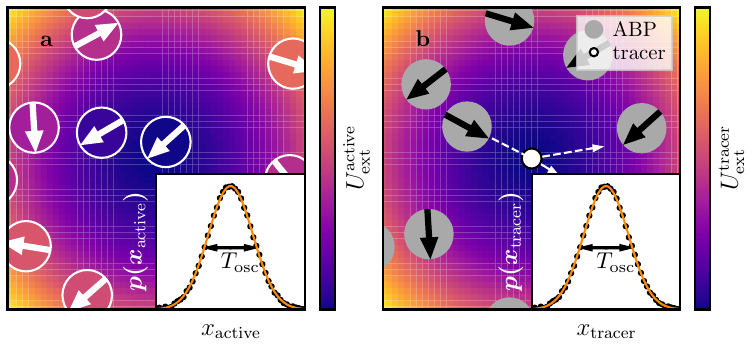}
		\caption{\textbf{Oscillator temperature.} \textbf{a} Schematic illustration of active particles in a harmonic potential $U_{\text{ext}}^{\text{active}}(r)=kr^2/2,~r=\sqrt{x^2+y^2}$ of strength $k$. The inset shows an exemplary distribution of the particle positions in $x$ direction $p(x_{\text{active}})$ from which $T_{\text{osc}}$ can be determined from the variance of a Gaussian that is fitted to the data. \textbf{b} Schematic visualization of a passive tracer particle (white) trapped in a harmonic potential $U_{\text{ext}}^{\text{tracer}}(r)=kr^2/2,~r=\sqrt{x^2+y^2}$ of strength $k$ and subject to a bath of non-trapped active particles (gray), which collide with the tracer particle (white arrows). The inset shows an exemplary distribution of the tracer position in $x$ direction $p(x_{\text{tracer}})$ from which $T_{\text{osc}}$ can be determined analogously as in panel a.}
		\label{fig:tosc-sketch}
	\end{figure}
	
	\item[(4)] \textbf{Oscillator temperature:} Let us now consider a particle that is confined in an external potential with a minimum at $\mathbf{r}=0$. In equilibrium, the position fluctuations $\langle\mathbf{r}^2\rangle$ are directly related to the thermodynamic temperature (Appendix D). For simplicity, let us consider a harmonic confinement, i.e., $U_{\text{ext}}(\mathbf{r})=k\mathbf{r}^2/2$.\cite{Caraglio_PhysRevLett_2022,Demery_JStatMech_2019,Dauchot_PhysRevLett_2019,Wexler_PhysRevRes_2020,Chaudhuri_JStatMechTheoExp_2021} For non-interacting particles in equilibrium, i.e., $v_0=0$ and $u=0$ in Eq.\ (\ref{eq:active-trans-leq}), one can show that $\langle r_i^2\rangle=\kb T/k$ with $i=x,y,z$. It is tempting to generalize this equilibrium result to define an oscillator temperature. Therefore, assuming that the system is isotropic, we define the oscillator temperature as \cite{DiBello_AnnPhys_2024,Wexler_PhysRevRes_2020}
	\begin{equation}
		\kb T_{\rm osc}\coloneqq k \erw{r_i^2},\,i=x,y,z. \label{eq:oscillator-temperature}
	\end{equation}
	There are two possibilities to measure the oscillator temperature: First, one can place the active particles themselves in the harmonic potential (Fig.\ \ref{fig:tosc-sketch}a). Note that in this scenario, for non-interacting particles in the limit of vanishing P\'eclet number (ideal gas), the oscillator temperature coincides with the virial temperature. Second, one can use a passive tracer particle trapped in a harmonic potential and interacting with surrounding non-trapped active particles (Fig.\ \ref{fig:tosc-sketch}b). The latter scenario is closely related to active heat engines, for which the definition in Eq.\ (\ref{eq:oscillator-temperature}) has been frequently used to map active heat engines onto an effective equilibrium system with a (time-dependent) effective temperature.\cite{Holubec_PhysRevE_2020,Guevara-Valadez_PhysicaA_2023,Holubec_PhysRevRes_2020,Wiese_ArXiv_2024} In terms of a general temperature definition, the tracer-based scenario has the drawback that the obtained temperature values depend on the mass of the tracer and its size, and defining a suitable tracer-based thermometer is only possible when choosing sufficiently small and heavy tracers.\cite{Hecht_PRL_2022,Loi_PhysRevE_2008} Furthermore, it has been shown that the position distribution of the tracer becomes non-Gaussian for certain $k$.\cite{Argun_PhysRevE_2016} Also in the former scenario, the strength $k$ of the harmonic potential has to be adjusted to the self-propulsion speed of the active particles such that they can still reach most positions inside the harmonic potential but cannot leave it across the periodic boundaries of the simulation box (see below). We remark that the dependence on the potential strength makes the use of this temperature questionable. Moreover, the use of the oscillator temperature causes problems when we consider interacting particles that repel each other and fill up the trapping potential from the center towards higher and higher potential values because it does not account for the contributions from the interactions between the particles. This inappropriately increases the value of the oscillator temperature and leads to a density-dependent temperature even in the equilibrium limit. Thus, we will conclude that the oscillator temperature is an unsuitable temperature definition. In contrast, the virial temperature consistently accounts for contributions from particle interactions [Eq.\ (\ref{eq:virial-temp-abps})], which is equal to the oscillator temperature in case of an ideal gas (see below). For simplicity, we only calculate $T_{\text{osc}}$ without using immersed tracer particles.
	
	For non-interacting ABPs (and AOUPs), the oscillator temperature has been calculated analytically and reads \cite{Caprini_JCP_2021}
	\begin{equation}
		\kb T_{\text{osc}} = \kb T_{\text{b}} + \frac{1 + \tau_{\text{p}}\gamma_{\text{t}}/m}{1 + \tau_{\text{p}}\gamma_{\text{t}}/m + \tau_{\text{p}}^2k/m}v_0^2\tau_{\text{p}}\gamma_{\text{t}}.\label{eq:Tosc-analytic}
	\end{equation}
	This expression is obtained from the AOUP model and coincides with the results for ABPs. It reflects the dependence of $T_{\text{osc}}$ on the strength $k$ of the harmonic potential and shows that it also depends on the ratio $\tau_{\text{p}}\gamma_{\text{t}}/m$. 
	
	\item[(5)] \textbf{Configurational temperature:} The configurational temperature provides another possibility to define temperature independently of the particle momenta. It can be derived from Eq.\ (\ref{eq:general-temp}) by choosing $\mathbf{B}(\boldsymbol{\Gamma})=-\boldsymbol{\nabla}U_{\text{tot}}(\lbrace \mathbf{r}_i\rbrace)$, where $U_{\text{tot}}(\lbrace \mathbf{r}_i\rbrace)$ denotes the total potential energy of the system. This yields \cite{Himpel_PhysRevE_2019,Han_JCP_2005}
	\begin{equation}
		k_{\text{B}}T_{\text{conf}}=\frac{\langle\boldsymbol{\nabla}U_{\text{tot}}\cdot\boldsymbol{\nabla}U_{\text{tot}}\rangle}{\langle\nabla^2U_{\text{tot}}\rangle}.\label{eq:Tconf}
	\end{equation}
	Here, $\boldsymbol{\nabla}$ is again the gradient operator in the $3N$-dimensional space. Recently, Saw et al.\ used the configurational temperature to measure the temperature of an active system.\cite{Saw_PhysRevE_2023,Saw_PhysRevE_2023_2} Note that for non-interacting active particles, $U_{\text{tot}}=U_{\text{ext}}$. In the special scenario of non-interacting particles in an external harmonic potential $U_{\text{ext}}(\mathbf{r})=k\mathbf{r}^2/2$, we get $T_{\rm conf}=T_{\rm osc}$.
	
	As shown in Fig.\ \ref{fig:tconf-sketch}, the configurational temperature measures how far a particle can ramp up the interaction potential. It is large if the forces ($\boldsymbol{\nabla}U_{\text{tot}}$) are large and if the curvature of the potential ($\nabla^2 U_{\text{tot}}$) is small. Therefore, contributions to $T_{\text{conf}}$ from particles residing near the minimum of the external potential and near the equilibrium distance for interacting particles are small, i.e., if all particles are placed in the potential minimum, $T_{\text{conf}}=0$ (Fig.\ \ref{fig:tconf-sketch}b). 
	
	\begin{figure}
		\centering
		\includegraphics[width=0.6\linewidth]{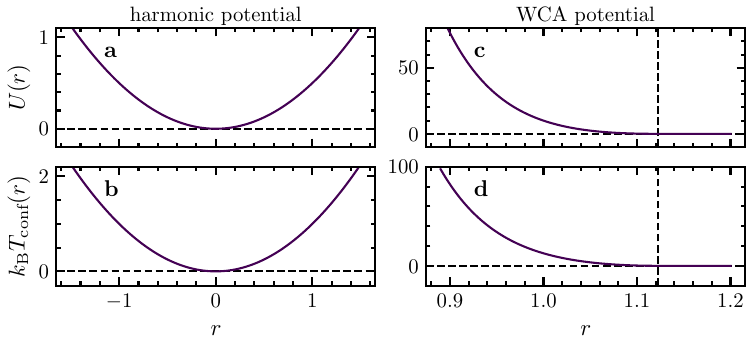}
		\caption{\textbf{Configurational temperature.} \textbf{a},\textbf{b} Exemplary harmonic potential $U(r)=0.5 k r^2$ and the corresponding contributions to the configurational temperature as defined in Eq.\ (\ref{eq:Tconf}) as function of the distance $r$, respectively. \textbf{c} Exemplary Weeks-Chandler-Anderson (WCA) potential as defined in Eq.\ (\ref{eq:wca}) and as used for the simulations in this work. \textbf{d} Corresponding contributions to the configurational temperature as defined in Eq.\ (\ref{eq:Tconf}) as function of the inter-particle distance $r$.}
		\label{fig:tconf-sketch}
	\end{figure}
	
\end{enumerate}

\subsection*{Dynamics-based definitions}
All previous possibilities to define temperature either exploit the velocities or the positions of the particles. However, one can also exploit dynamical properties of active systems to define temperature. In particular, we present two approaches, one based on the Einstein relation and one following Cugliandolo and Kurchan based on linear response theory.\cite{Cugliandolo_FluctNoiseLett_2019,Cugliandolo_PhysRevE_1997}

\begin{enumerate}
	\item[(6)] \textbf{Einstein temperature:} Let us again consider free non-interacting passive Brownian particles, i.e., $v_0=0$, $u=0$, and $\mathbf{F}_{\text{ext}}=0$ in Eq.\ (\ref{eq:active-trans-leq}). Then, the diffusion coefficient $D$ is connected to the bath temperature via the Einstein relation: $D=\kb T_{\text{b}}/\gamma_{\text{t}}$.\cite{Einstein_AnnPhys_1905} We can now define a temperature for active particles based on their effective long-time diffusion coefficient $D_{\text{eff}}$, which can be calculated from the mean-square displacement (MSD) of the active particles (Fig.\ \ref{fig:tein-sketch}), by exploiting the Einstein relation. In particular, we define the Einstein temperature $T_{\text{Ein}}$ as \cite{Szamel_EPL_2017}
	\begin{equation}
		\kb T_{\text{Ein}}\coloneqq\gamma_{\text{t}} D_{\text{eff}}. \label{eq:einstein-temperature}
	\end{equation}
	If the active particles interact with each other, i.e., $u\neq 0$ in Eq.\ (\ref{eq:active-trans-leq}), $\gamma_{\text{t}}$ has to be replaced by an effective drag coefficient $\gamma_{\text{eff}}$ that is calculated from the response of a tracer particle to a constant force $\mathbf{F}=F\mathbf{e}_x$ in the presence of the considered active system, i.e.,
	\begin{equation}
		\gamma_{\text{eff}} = F/\Lim{t}{\infty}\erw{v_x(t)}. \label{eq:gamma-calc}
	\end{equation}
	Note that for sufficiently low density, we have $\gamma_{\text{eff}}\approx \gamma_{\text{t}}$.
	
	For free non-interacting inertial ABPs, the effective diffusion coefficient has been calculated analytically and reads \cite{Sandoval_PhysRevE_2020,Sprenger_PhysRevE_2021,Scholz_NatCom_2018}
	\begin{equation}
		D_{\text{eff}}^{\text{ABP}} = D_{\text{t}} + \frac{v_0^2\tau_{\text{p}}}{2}e^{S_{\text{R}}} S_{\text{R}}^{1-S_{\text{R}}}\Gamma(S_{\text{R}},0,S_{\text{R}}),\label{eq:D-analytic-ABP}
	\end{equation}
	where $D_{\text{t}}=\kb T_{\text{b}}/\gamma_{\rm t}$ denotes the translational diffusion coefficient, $S_{\text{R}}=\frac{I}{\tau_{\text{p}}\gamma_{\text{r}}}$, and $\Gamma(a,b,c)=\int_{b}^{c}\text{d}q\,q^{a-1}e^{-q}$. Here, $I$ denotes the moment of inertia of the active particles. In the overdamped limit $S_{\text{R}}\rightarrow 0$, we obtain the following popular result for the active diffusion coefficient: $D_{\text{eff}}=D_{\text{t}}+v_0^2\tau_{\rm p}/2$,\cite{Howse_PhysRevLett_2007,TenHagen_JPhysCondensMatter_2011,Hecht_ArXiv_2021} yielding $\kb T_{\text{Ein}}=\kb T_{\text{b}}+\gamma_{\text{t}}v_0^2\tau_{\text{p}}/2$.
	
	\begin{figure}
		\centering
		\includegraphics[width=0.5\linewidth]{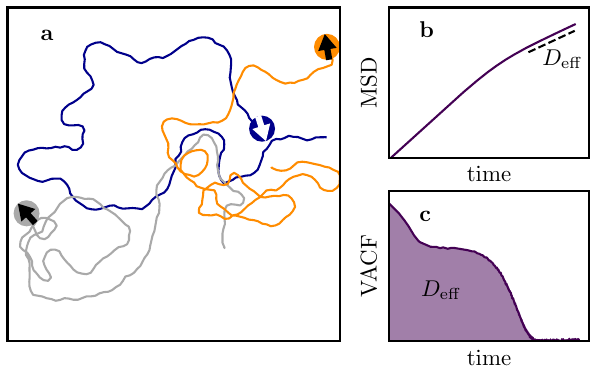}
		\caption{\textbf{Einstein temperature.} \textbf{a} Exemplary trajectories of free non-interacting ABPs. Arrows denote the self-propulsion directions. \textbf{b} Mean-square displaced (MSD) of the active particles and long-time effective diffusion coefficient $D_{\text{eff}}$ yielding the Einstein temperature as defined in Eq.\ (\ref{eq:einstein-temperature}). \textbf{c} Alternative calculation of the long-time diffusion coefficient $D_{\text{eff}}$ from the integral of the velocity auto-correlation function (VACF) of the active particles.}
		\label{fig:tein-sketch}
	\end{figure}
	
	\item[(7)] \textbf{Effective temperature:} Inspired from glassy systems \cite{Berthier_JCP_2002,Berthier_PhysRevLett_2002} and following Refs.\ \onlinecite{Cugliandolo_FluctNoiseLett_2019,Petrelli_PhysRevE_2020,Loi_SoftMatter_2011,Cugliandolo_JPhysAMathTheor_2011}, we now define the so-called effective temperature of an active system. This expression is inspired by linear response theory and is defined as the ratio between the mean-square displacement ${\rm MSD}(t)=\erw{\left[\mathbf{r}(t)-\mathbf{r}(0)\right]^2}$ and the susceptibility $\chi$ in the long-time limit, i.e.,  
	\begin{equation}
		\kb T_{\rm eff}(t) \coloneqq \lim\limits_{t\gg 1}\frac{\text{MSD}(t)}{2d\chi(t)},\label{eq:Teff}
	\end{equation}
	where $d$ is the number of spatial dimensions. To calculate the susceptibility, one can use the Malliavin weights sampling (MWS), as used in Refs.\ \onlinecite{Petrelli_PhysRevE_2020,Cugliandolo_FluctNoiseLett_2019}, or approaches that are based on the simulation of a perturbed and an unperturbed system with the same noise realizations (Fig.\ \ref{fig:teff-sketch}).\cite{Villamaina_JStatMech_2008,Fodor_PhysRevLett_2016,Ciccotti_JStatPhys_1979} Here, we will use the latter approach to numerically determine $T_{\text{eff}}$ (see Appendix E for details). Note that the calculation of the effective temperature requires to average over many independent ensembles, which is computationally expensive. In particular, we average over 100 independent simulation runs for each data point and over time in the diffusive long-time regime of the MSD. Due to the high computational costs, we calculated the effective temperature exemplarily for some parameter regimes, and we only consider data points with reasonably good statistics, i.e., data points with a standard deviation that is at least smaller than the value itself.
	
	\begin{figure}
		\centering
		\includegraphics[width=0.7\linewidth]{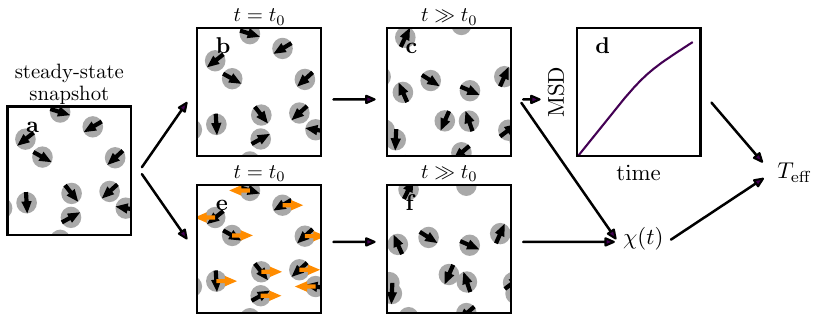}
		\caption{\textbf{Effective temperature.} Schematic visualization of the calculation of the effective temperature $T_{\text{eff}}$ as defined in Eq.\ (\ref{eq:Teff}). Starting from a snapshot of the system of active particles in the steady state (panel \textbf{a}), a copy of the system is created and perturbed by a small perturbing force at a fixed time $t=t_0$ (orange arrows in panel \textbf{e}). The original system (panel \textbf{b}) and the perturbed system are then simulated with the same noise realizations up to a time $t\gg t_0$ (panels \textbf{c} and \textbf{f}). From the unperturbed system, the mean-square displacement (MSD, panel \textbf{d}) is calculated and from the comparison of both systems, the susceptibility $\chi(t)$ is obtained. Finally, this leads to $T_{\text{eff}}$ following Eq.\ (\ref{eq:Teff}).}
		\label{fig:teff-sketch}
	\end{figure}
	
\end{enumerate}

%%%%%%%%%%%%%%%%%%%%%%%%%%%%%%%%%%%%%%%%%%%%%%%%%%%%%%%%%%%%%%%%%%%%%%%%%%%%%%%%%%%%%%%%%%%%%%%%%%%%%%%%%%%%%
% NUMERICAL RESULTS
%%%%%%%%%%%%%%%%%%%%%%%%%%%%%%%%%%%%%%%%%%%%%%%%%%%%%%%%%%%%%%%%%%%%%%%%%%%%%%%%%%%%%%%%%%%%%%%%%%%%%%%%%%%%%
\section*{Simulation Results}
\subsection*{Brownian dynamics simulations of the ABP model}
To systematically compare the introduced possibilities to define temperature, we perform Brownian dynamics simulations of systems of $N=2\times10^{4}$ inertial ABPs as described by Eqs.\ (\ref{eq:active-trans-leq}) and (\ref{eq:abp-rot-leq}). The interaction between the ABPs is modeled by the Weeks-Chandler-Anderson (WCA) potential \cite{Weeks_JCP_1971}
\begin{equation}
	u(r_{ij})=
	\begin{cases}
		4\epsilon\left[\left(\frac{\sigma}{r_{ij}}\right)^{12}-\left(\frac{\sigma}{r_{ij}}\right)^{6}\right]+\epsilon,~&r_{ij}/\sigma\leq 2^{1/6} \\ 0, ~&\text{else}
	\end{cases}\label{eq:wca}
\end{equation}
with particle diameter $\sigma$, strength $\epsilon$, and $r_{ij}=\left\lvert\mathbf{r}_i-\mathbf{r}_j\right\rvert$. We use the particle diameter $\sigma$ as length unit, the persistence time $\tau_{\text{p}}=1/D_{\text{r}}$ as time unit, and $\kb T_{\text{b}}$ as energy unit. Here, $D_{\text{r}}=k_{\rm B} T_{\text{b}}/\gamma_{\text{r}}$ denotes the rotational diffusion coefficient and $T_{\text{b}}$ is the bath temperature. We fix the interaction strength to $\epsilon/(k_{\rm B}T_{\rm b})=10$ or to $\epsilon=0$ for non-interacting particles. For simplicity, we choose $\gamma_{\text{t}}=\gamma_{\text{r}}/\sigma^2$. As free parameters, we vary the dimensionless mass $M=m/(\gamma_{\text{t}}\tau_{\rm p})=\tau_{\text{m}}/\tau_{\text{p}}$ (and accordingly, the moment of inertia $I$), the P\'eclet number $\text{Pe}=v_0/\sqrt{2D_{\text{r}}D_{\text{t}}}$ (which quantifies the strength of self-propulsion relative to diffusive motion), and the total packing fraction $\varphi_{\text{tot}}=N\pi\sigma^2/(4A)$, where $A=L^2$ is the area of the two-dimensional quadratic simulation box of box length $L$. Here, $D_{\text{t}}=\kb T_{\text{b}}/\gamma_{\text{t}}$ is the translational diffusion coefficient and $\tau_{\text{m}}=m/\gamma_{\text{t}}$ is the inertial time scale. The Langevin equations [Eqs.\ (\ref{eq:active-trans-leq}) and (\ref{eq:abp-rot-leq})] are solved numerically with LAMMPS \cite{Thompson_CompPhysComm_2022} using a time step $\Delta t/\tau_{\rm p}=10^{-5}$ and periodic boundary conditions. We run the simulations first for a time of $200\,\tau_{\rm p}$ to reach a steady state and afterwards for a time of $800\,\tau_{\rm p}$ for computing time averages of observables in the steady state. For simulations in a harmonic confinement, we have chosen $k\propto 2\gamma_{\text{t}}v_0/L$ such that $|\mathbf{F}_{\text{ext}}(L/2)|=\gamma_{\text{t}}v_0$ to ensure that the active particles are able to reach each position in the harmonic potential but cannot leave it across the periodic boundaries of the simulation box. All simulation data has been analyzed with Python using the recently developed active matter evaluation package (AMEP).\cite{Hecht_ArXiv_2024}

\subsection*{P\'eclet dependence}
\begin{figure}
	\centering
	\includegraphics[width=1.0\linewidth]{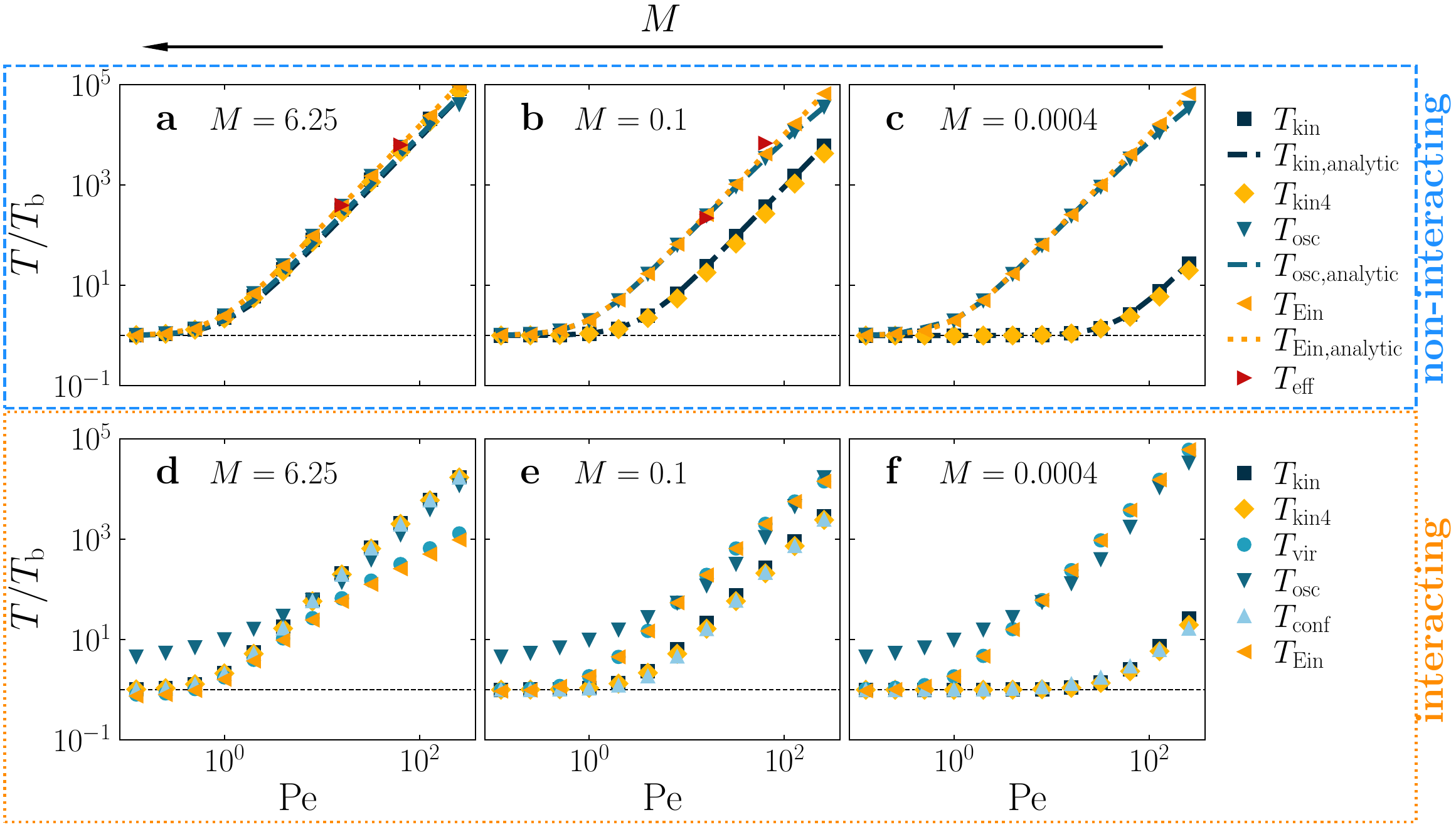}
	\caption{\textbf{Pe dependence.} Temperature as function of P\'eclet number Pe for three different masses $M=m/(\gamma_{\text{t}}\tau_{\text{p}})$ as given in the key. \textbf{a--c} Results for non-interacting ABPs and \textbf{d--f} for interacting ABPs at total packing fraction of $\varphi_{\rm tot}=0.025$. For the non-interacting case, analytical expressions are shown for the kinetic temperature [$T_{\text{kin,analytic}}$, Eqs.\ (\ref{eq:Tkin-analytic-free}) and (\ref{eq:Tkin-alpha-free})], for the oscillator temperature [$T_{\text{osc,analytic}}$, Eq.\ (\ref{eq:Tosc-analytic})], and for the Einstein temperature [$T_{\text{Ein,analytic}}$, Eqs.\ (\ref{eq:einstein-temperature}) and (\ref{eq:D-analytic-ABP})]. The black dashed line denotes the bath temperature $T_{\text{b}}$.}
	\label{fig:fig2}
\end{figure}

Let us first discuss the Pe dependence of the considered temperatures. For simplicity, we only consider parameter regimes in which the system does not phase separate \cite{Digregorio_PhysRevLett_2018} and study two scenarios: non-interacting ABPs and interacting ABPs at total packing fraction $\varphi_{\text{tot}}=0.025$. To scan all regimes from the near-equilibrium case to the strongly active regime, we vary Pe from $\text{Pe}=0.125$ to $\text{Pe}=256$. To also explore different regimes from the strongly inertial regime to the overdamped regime, we determine the temperature for three different masses $M\in \lbrace 0.0004, 0.1, 6.25\rbrace$. The results are shown in Fig.\ \ref{fig:fig2} together with the corresponding analytical expressions for non-interacting ABPs as discussed above. These expressions perfectly match with the numerical results obtained from the Brownian dynamics simulations (Fig.\ \ref{fig:fig2}a--c). As expected, all temperatures increase with increasing Pe (Fig.\ \ref{fig:fig2}). For very low Pe, the system essentially behaves as an equilibrium system and all temperatures coincide (except for the oscillator temperature for interacting ABPs as we discuss in more detail below).

\paragraph*{Non-interacting active particles}
Let us now first focus on the non-interacting case. Here, for large mass $M$ (Fig.\ \ref{fig:fig2}a), all temperatures lead to the same value and the curves collapse to one master curve. In this case, the persistence time $\tau_{\text{p}}$ is small compared to the inertial time $\tau_{\text{m}}$, which leads to a vanishing entropy production rate such that the system approaches an effective equilibrium state.\cite{Omar_JCP_2023,Caprini_JCP_2023,Caprini_JCP_2021} When decreasing the mass, i.e., for $\tau_{\text{p}}/\tau_{\text{m}}\gg 1$, different temperatures generally lead to different values (Fig.\ \ref{fig:fig2}b,c). Notably, the two kinetic temperatures $T_{\text{kin}}$ and $T_{\text{kin4}}$ lead to very similar temperature values suggesting that the velocity distributions are approximately Gaussian. Remarkably, also $T_{\text{osc}}$ and $T_{\text{Ein}}$ lead to very similar temperature values even for very large Pe, where they significantly differ from $T_{\text{kin}}$ and $T_{\text{kin4}}$. In fact, we find $T_{\text{osc}}\geq T_{\text{kin}}$ in accordance with previous literature.\cite{Wexler_PhysRevRes_2020} Note that $T_{\text{osc}}$ and $T_{\text{kin}}$ might coincide after doing a force renormalization as demonstrated in Ref.\ \onlinecite{Shea_SoftMatter_2024}. The difference to the kinetic temperatures further increases when decreasing the mass of the active particles (Fig.\ \ref{fig:fig2}b,c) indicating that velocity-based definitions strongly depend on the dimensionless particle mass while $T_{\text{osc}}$ and $T_{\text{Ein}}$ do not as we shall see. The effective temperature $T_{\text{eff}}$ is similar to $T_{\text{osc}}$ and $T_{\text{Ein}}$ (Fig.\ \ref{fig:fig2}a,b). Note that we show only a few data points for $T_{\text{eff}}$ because its computation is numerically rather costly.

\paragraph*{Interacting active particles}
If we now consider interactions between the active particles, we qualitatively obtain the same results. Again, all temperature values obtained with the different possibilities to define temperature except the oscillator temperature coincide for large masses and lead to the bath temperature at small Pe (Fig.\ \ref{fig:fig2}d). However, the oscillator temperature $T_{\text{osc}}$ saturates at a temperature larger than the bath temperature. This is because (i) the confining potential pushes the particles together such that the particles may form a dense cluster around the minimum of the confining potential and (ii) not all particles can be placed in the potential minimum in the initial state of the simulation. The latter adds some additional potential energy to the particles. As a consequence, particles at the border of the cluster have a large potential energy and lead to a large contribution to the position fluctuations $\langle r^2\rangle$. Hence, the oscillator temperature can reach values higher than the bath temperature even for passive particles. In particular, as mentioned earlier, the oscillator temperature does not appropriately consider contributions coming from the interaction forces between the particles. In turn, the virial temperature systematically considers these contributions and therefore leads to correct temperature values. Hence, the oscillator temperature is considered as unsuitable to measure temperature. At large Pe, another deviation is visible: The Einstein temperature $T_{\text{Ein}}$ and the virial temperature $T_{\text{vir}}$ lead to smaller temperature values at large Pe compared to the other temperature values especially in the case of large particle mass (Fig.\ \ref{fig:fig2}d). This is because collisions slow down the particles and lead to a smaller diffusion coefficient. This effect is stronger at large Pe due to a higher collision rate and it is also stronger for heavier active particles because they need a comparatively long time to reach their terminal speed after each collision. However, similar to the non-interacting scenario, the Einstein temperature $T_{\text{Ein}}$ and the oscillator temperature $T_{\text{osc}}$ almost coincide at intermediate and small masses and for large Pe (Fig.\ \ref{fig:fig2}d--f). This is because they both effectively measure position fluctuations and therefore, approximately coincide. Note that $T_{\text{Ein}}$, $T_{\text{vir}}$, and $T_{\text{osc}}$ do not coincide with the other temperatures because they have a weaker mass dependence as we will discuss further below. Remarkably, the configurational temperature $T_{\text{conf}}$ coincides with $T_{\text{kin}}$ and $T_{\text{kin4}}$ for all parameters (Fig.\ \ref{fig:fig2}d--f). It measures how far an active particle can ramp up the interaction potential and therefore, it is directly related to the kinetic energy of the particles that is converted into potential energy during collisions for example. Hence, $T_{\text{conf}}$ leads to very similar temperature values as $T_{\text{kin}}$ and $T_{\text{kin4}}$. 

\begin{figure}
	\centering
	\includegraphics[width=1.0\linewidth]{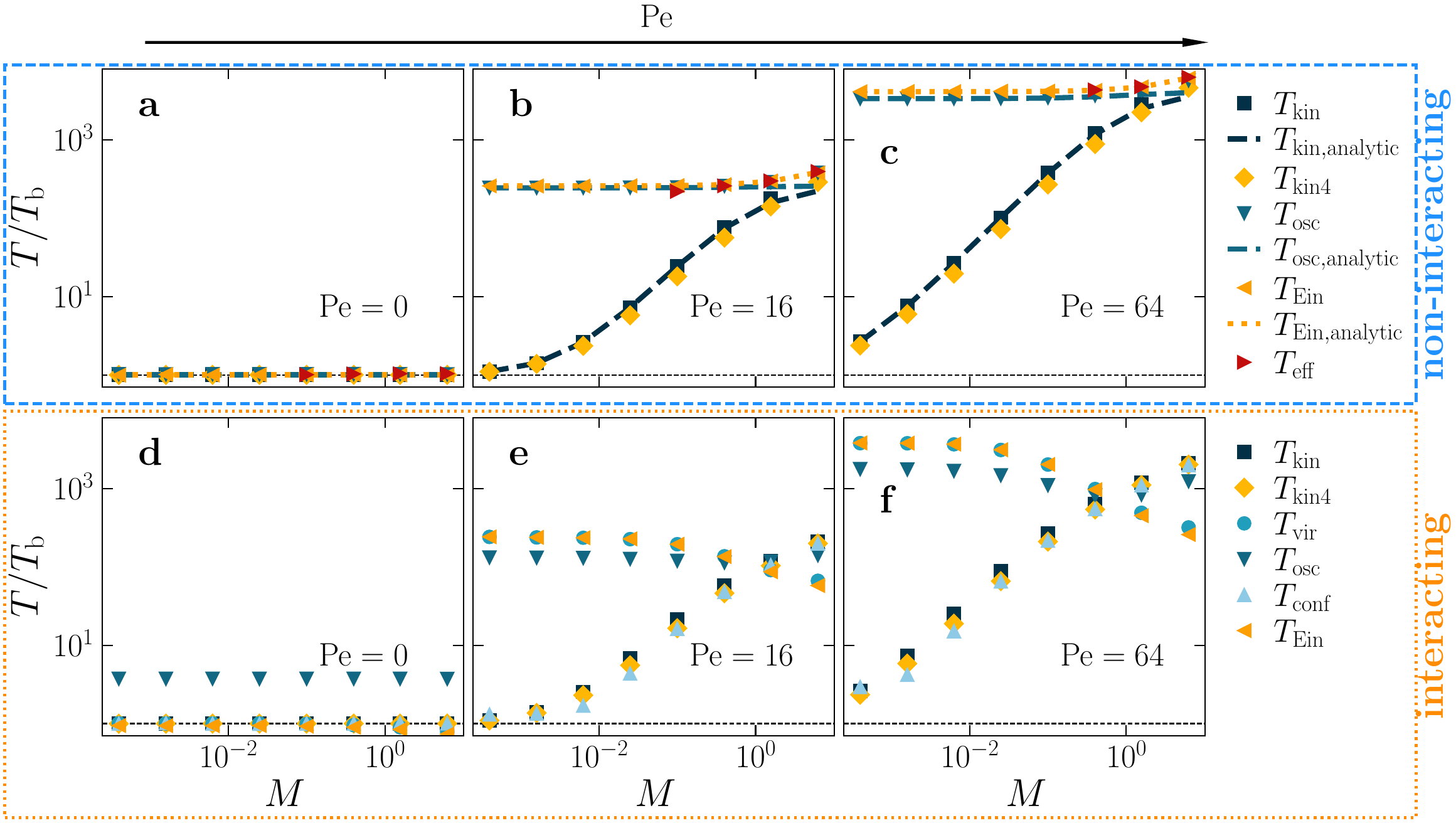}
	\caption{\textbf{Mass dependence.} Temperature as a function of the particle mass $M=m/(\gamma_{\text{t}}\tau_{\text{p}})$ for three different values of Pe as given in the key. \textbf{a--c} Results for non-interacting ABPs and \textbf{d--f} for interacting ABPs at total packing fraction of $\varphi_{\rm tot}=0.025$. For the non-interacting case, analytical expressions are shown for the kinetic temperature [$T_{\text{kin,analytic}}$, Eqs.\ (\ref{eq:Tkin-analytic-free}) and (\ref{eq:Tkin-alpha-free})], for the oscillator temperature [$T_{\text{osc,analytic}}$, Eq.\ (\ref{eq:Tosc-analytic})], and for the Einstein temperature [$T_{\text{Ein,analytic}}$, Eqs.\ (\ref{eq:einstein-temperature}) and (\ref{eq:D-analytic-ABP})]. The black dashed line denotes the bath temperature $T_{\text{b}}$.}
	\label{fig:fig1}
\end{figure}

\subsection*{Mass dependence}
To obtain further insights into the parameter dependencies of the different temperatures, we now analyze the mass dependence in more detail. We vary $M=m/(\gamma_{\text{t}}\tau_{\text{p}})$ from $0.0004$ to $6.25$ for $\text{Pe}\in\lbrace 0, 16, 64\rbrace$. The results are shown in Fig.\ \ref{fig:fig1} again together with the analytical results for non-interacting ABPs. For the latter, the numerical results perfectly coincide with analytical expressions discussed above (Fig.\ \ref{fig:fig1}a--c). As expected from equilibrium thermodynamics, all temperatures lead to the same temperature values, namely the bath temperature $T_{\text{b}}$, for $\text{Pe}=0$. For $\text{Pe}> 0$, the different temperatures again lead to different temperature values and exhibit an important mass dependence: While the oscillator temperature $T_{\text{osc}}$ does not depend on $M$ and the Einstein temperature, the effective temperature, and the virial temperature only show a weak mass dependence, both kinetic temperatures $T_{\text{kin}}$ and $T_{\text{kin4}}$ as well as the configurational temperature $T_{\text{conf}}$ feature a strong mass dependence (Fig.\ \ref{fig:fig1}). In the absence of interactions, this is because the particles move with their terminal self-propulsion speed $\langle |\mathbf{v}|\rangle\approx v_0$, and accordingly, we have $T_{\text{kin}}\approx mv_0^2/2\propto m$ for large Pe. Thus, for $m\rightarrow 0$ (i.e, $M\rightarrow 0$), the active contribution to $T_{\text{kin}}$ vanishes and we have $T_{\text{kin}}\approx T_{\text{b}}$. In turn, $T_{\text{Ein}}$, $T_{\text{eff}}$, $T_{\text{vir}}$, and $T_{\text{osc}}$ only slightly depend on $M$ (Fig.\ \ref{fig:fig1}b,c): Their calculation is based on position fluctuations which depend only weakly on $M$ in some parameter regimes if the particles (on average) move with their terminal self-propulsion speed $v_0$. The observed trends are robust and still apply in the presence of interactions (Fig.\ \ref{fig:fig1}d--f).

\begin{figure}
	\centering
	\includegraphics[width=0.65\linewidth]{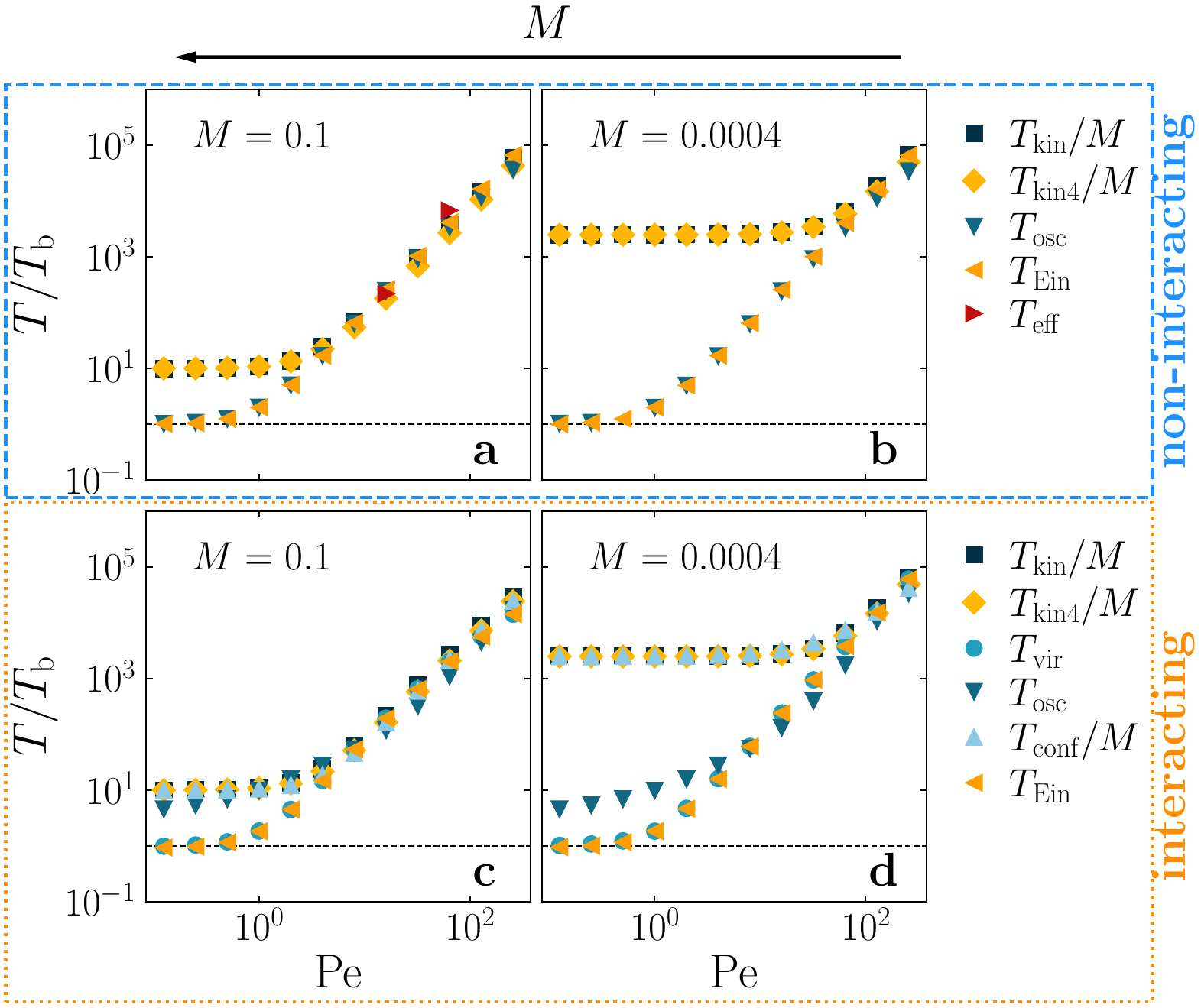}
	\caption{\textbf{Pe dependence of rescaled temperatures.} Temperature as a function of Pe for two different masses $M=m/(\gamma_{\text{t}}\tau_{\text{p}})$ as given in the key. All temperatures with a strong mass dependence are rescaled with $M$. \textbf{a},\textbf{b} Results for non-interacting ABPs, i.e., $u=0$ in Eq.\ (\ref{eq:active-trans-leq}). \textbf{c},\textbf{d} Results for interacting ABPs at total packing fraction $\varphi_{\text{tot}}=0.025$. The black dashed line denotes the bath temperature $T_{\text{b}}$.}
	\label{fig:fig2-rescaled}
\end{figure}

\begin{figure}
	\centering
	\includegraphics[width=0.65\linewidth]{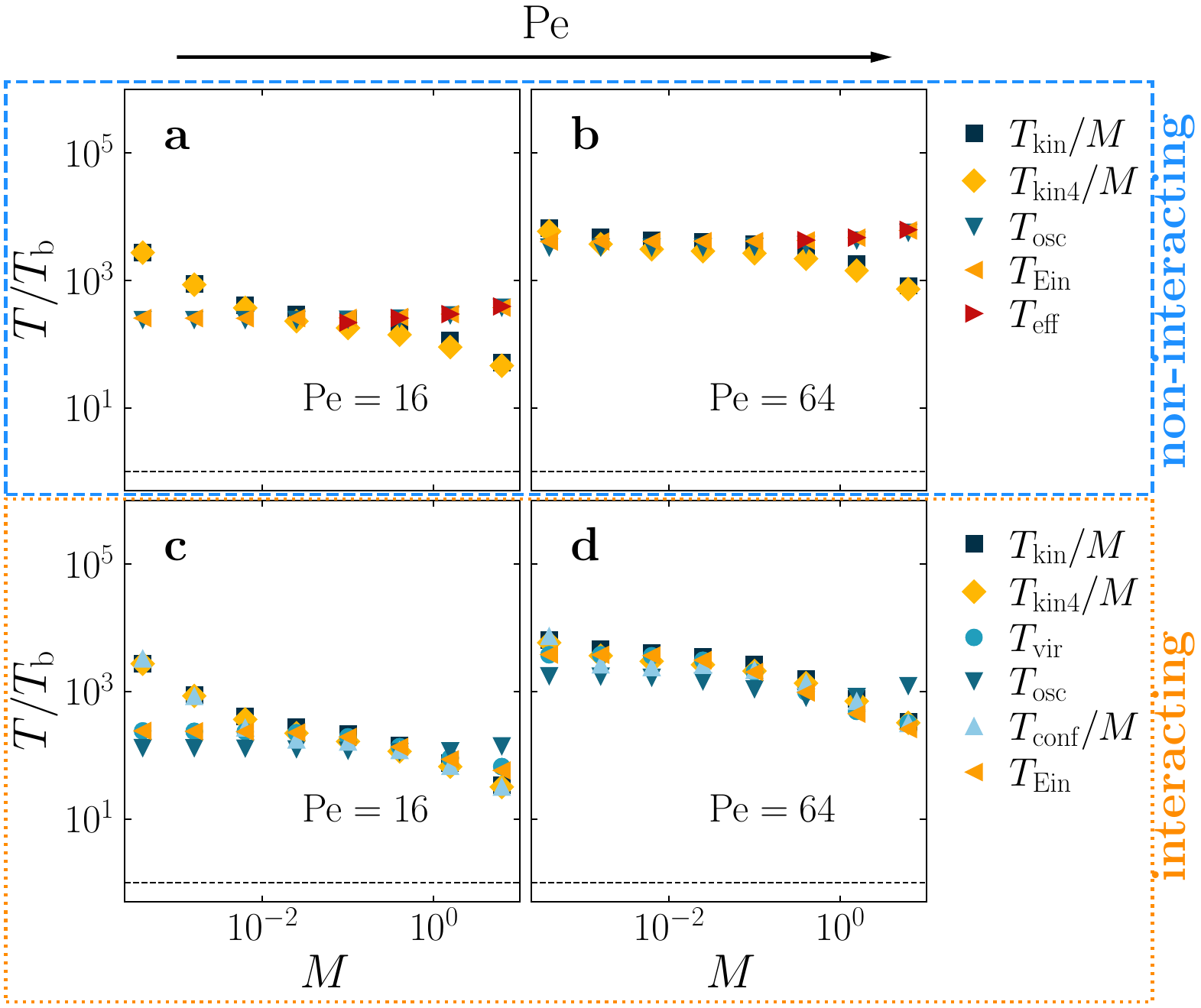}
	\caption{\textbf{Mass dependence of rescaled temperatures.} Temperature as a function of the dimensionless particle mass $M=m/(\gamma_{\text{t}}\tau_{\text{p}})$ for two different Pe as given in the key. All temperatures with a strong mass dependence are rescaled with $M$. \textbf{a},\textbf{b} Results for non-interacting ABPs, i.e., $u=0$ in Eq.\ (\ref{eq:active-trans-leq}). \textbf{c},\textbf{d} Results for interacting ABPs at total packing fraction $\varphi_{\text{tot}}=0.025$. The black dashed line denotes the bath temperature $T_{\text{b}}$.}
	\label{fig:fig1-rescaled}
\end{figure}

\paragraph*{Mass scaling}
From Fig.\ \ref{fig:fig1}, we see that some temperature definitions strongly depend on the mass of the active particles. Inspired by the proportionality to $m$ of the kinetic temperatures for large Pe, where $\langle \mathbf{v}\rangle \approx v_0$ [Eqs.\ (\ref{eq:kinetic-temperature}) and (\ref{eq:fourth-kinetic-temperature})], we divide all temperatures that show a strong mass dependence ($T_{\text{kin}}$, $T_{\text{kin4}}$, and $T_{\text{conf}}$) by $M=m/(\gamma_{\text{t}}\tau_{\text{p}})$ (Figs.\ \ref{fig:fig2-rescaled} and \ref{fig:fig1-rescaled}). From the Pe-dependence, we see that now all definitions lead to similar temperatures at large Pe for both non-interacting and interacting ABPs (Fig.\ \ref{fig:fig2-rescaled}). Note that the regime in which the (rescaled) temperatures coincide is larger if the active particles are heavier. This is reflected by the analytical expressions discussed above, which show that for low or moderate Pe, the kinetic temperature is not simply proportional to $m$ but has a more complicated mass dependence encoded in the factor $\alpha$ for example [cf.\ Eq.\ (\ref{eq:Tkin-alpha-free})]. This becomes visible in Fig.\ \ref{fig:fig1-rescaled}, which reflects that the different temperatures do of course not fully match even if rescaled with $M$.

\subsection*{Effect of the packing fraction}
\begin{figure*}
	\centering
	\includegraphics[width=1.0\linewidth]{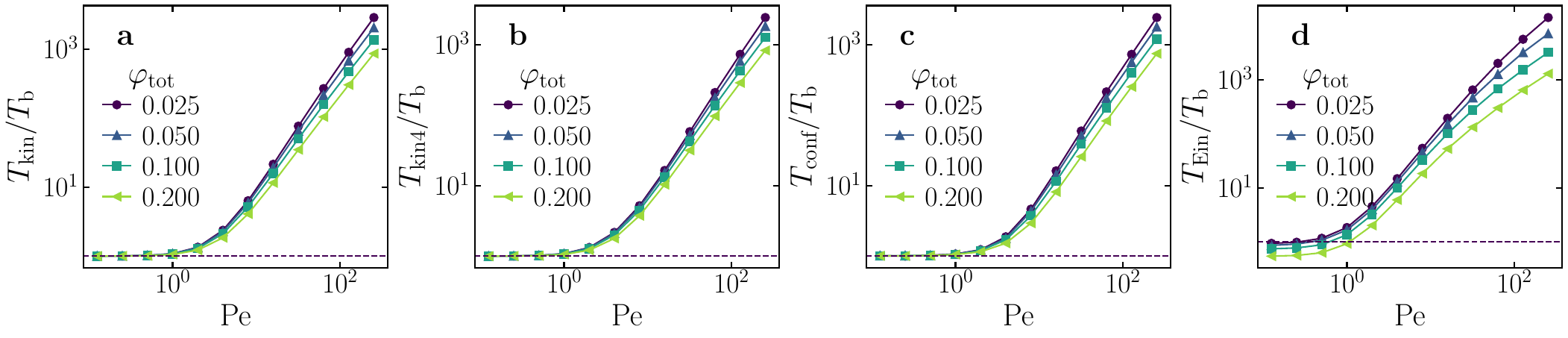}
	\caption{\textbf{Effect of the packing fraction.} Temperature values for a fixed mass $M=0.1$ as function of the P\'eclet number Pe and at different total packing fractions $\varphi_{\text{tot}}$ (indicated in the keys) for \textbf{a} the kinetic temperature $T_{\text{kin}}$, \textbf{b} the fourth-moment kinetic temperature $T_{\text{kin4}}$, \textbf{c} the configurational temperature $T_{\text{conf}}$, and \textbf{d} the Einstein temperature $T_{\text{Ein}}$. The black dashed line denotes the bath temperature $T_{\text{b}}$.}
	\label{fig:fig-11}
\end{figure*}

Finally, we exemplarily analyzed the effect of the total packing fraction on the values of the kinetic temperatures $T_{\text{kin}}$ and $T_{\text{kin4}}$, the configurational temperature $T_{\text{conf}}$, and the Einstein temperature $T_{\text{Ein}}$. We have chosen the total packing fraction as $\varphi_{\text{tot}}\in\lbrace 0.025, 0.05, 0.1, 0.2\rbrace$ such that the system is still uniform and does not undergo MIPS \cite{Digregorio_PhysRevLett_2018,Dai_SoftMatter_2020,Hecht_PRL_2022} and fixed an intermediate mass of $M=0.1$. As shown in Fig.\ \ref{fig:fig-11}, increasing the packing fraction decreases the temperature values but the overall Pe dependence is the same. This is because increasing the total packing fraction leads to an increased collision rate. The collisions tend to slow down the active particles and hinder the particles to reach their self-propulsion speed $v_0$. Hence, increasing the packing fraction opposes the effect of activity on the average speed of the particles and reduces the values of the considered temperatures. Close to equilibrium, all temperatures are equal to the bath temperature for all packing fractions except for the Einstein temperature (Fig.\ \ref{fig:fig-11}d). Here, we have used the same drag coefficient $\gamma_{\text{t}}$ for all packing fractions to calculate $T_{\text{Ein}}$. However, increasing $\varphi_{\text{tot}}$ reduces the MSD, and hence, also $T_{\text{Ein}}$. In particular, the resulting diffusion coefficient becomes smaller for larger packing fraction. Hence, also $T_{\text{Ein}}$ decreases with increasing $\varphi_{\text{tot}}$. This also happens close to equilibrium and leads to an Einstein temperature slightly smaller than the bath temperature. One could compensate for this effect by calculating the effective drag coefficient as shown in Eq.\ (\ref{eq:gamma-calc}).

%%%%%%%%%%%%%%%%%%%%%%%%%%%%%%%%%%%%%%%%%%%%%%%%%%%%%%%%%%%%%%%%%%%%%%%%%%%%%%%%%%%%%%%%%%%%%%%%%%%%%%%%%%%%%
%%%%%%%%%%%%%%%%%%%%%%%%%%%%%%%%%%%%%%%%%%%%%%%%%%%%%%%%%%%%%%%%%%%%%%%%%%%%%%%%%%%%%%%%%%%%%%%%%%%%%%%%%%%%%
% SUMMARY AND OUTLOOK
%%%%%%%%%%%%%%%%%%%%%%%%%%%%%%%%%%%%%%%%%%%%%%%%%%%%%%%%%%%%%%%%%%%%%%%%%%%%%%%%%%%%%%%%%%%%%%%%%%%%%%%%%%%%%
%%%%%%%%%%%%%%%%%%%%%%%%%%%%%%%%%%%%%%%%%%%%%%%%%%%%%%%%%%%%%%%%%%%%%%%%%%%%%%%%%%%%%%%%%%%%%%%%%%%%%%%%%%%%%
\section*{Conclusions}
Our analytical and numerical results show that different possibilities to define temperature typically lead to different temperature values. However, close to equilibrium, all temperatures coincide. In active systems, one can approach (effective) equilibrium states in two ways: First, in the limit $\text{Pe}\rightarrow 0$, activity vanishes and the system forms an equilibrium system made of passive Brownian particles, for which all temperatures coincide with the bath temperature. Second, in the limit $M\rightarrow\infty$, the persistence time $\tau_{\text{p}}$ becomes small compared to the inertial time $\tau_{\text{m}}=m/\gamma_{\text{t}}$. Then, the motion of the active particles is dominated by (rotational) diffusion and the system reaches an effective equilibrium state at a larger temperature than the bath temperature. This is also indicated by a vanishing entropy production rate in the limit $1/M=\tau_{\text{p}}/\tau_{\text{m}}\rightarrow 0$.\cite{Caprini_JCP_2023,Fodor_PhysRevLett_2016}

It is now tempting to distinguish between ``good'' and ``bad'' temperature definitions: A ``good'' temperature definition should provide consistent temperature values that are independent of details of the thermometer and the confining potential. Therefore, we conclude that the oscillator temperature and any tracer-based temperature definition generally can be considered as a comparatively ``bad'' definition of temperature in active systems. This is because the oscillator temperature strongly depends on the potential strength $k$ and bears the risk of not agreeing with the bath temperature in the equilibrium limit for interacting particles because it does not appropriately account for interaction forces compared to the virial temperature for example. In turn, any tracer-based definition requires heavy and small tracer particles such that they (i) follow the slow dynamics of the active system and (ii) do not affect its structure.\cite{Hecht_PRL_2022,Loi_PhysRevE_2008} Furthermore, some temperatures are computationally demanding such as the effective temperature. In contrast, all other temperatures can be considered as comparatively ``good'' in the sense that they do not suffer from these drawbacks. While their values of course depend on details of the considered system (dimensionless parameters such as the reduced mass, P\'eclet number, and packing fraction), we found that several temperatures approximately coincide even far from equilibrium. Concretely, the kinetic temperature $T_{\text{kin}}$, the fourth-moment-based kinetic temperature $T_{\text{kin4}}$, and the configurational temperature $T_{\text{conf}}$ constitute a class of temperatures that all assume very similar temperature values over a wide parameter range. Notably, the virial temperature $T_{\text{vir}}$, the Einstein temperature $T_{\text{Ein}}$, the oscillator temperature $T_{\text{osc}}$, and the effective temperature $T_{\text{eff}}$ form a second class of temperatures whose values approximately coincide with each other but which strongly differ from those of the first class. Beyond that, we found that the two different classes of temperatures can be matched in the far-from equilibrium regime where the system is dominated by activity (large Pe, small mass) by rescaling temperatures with the particle mass. 

Overall, regarding the question ``How to define temperature in active systems?'', we note that our numerical results reflect the general expectation that far from equilibrium, different temperatures lead to different temperature values. This is because the particle positions and velocities are non-trivially coupled in active systems, and in general, often follow different non-Boltzmann distributions. This implies that it is impossible to uniquely quantify fluctuations in active systems based on a single temperature parameter. However, beyond this generic fact, we found that certain possibilities to define temperature are advantageous over others in the sense that they are (i) easy to calculate from numerical (or experimental) data, (ii) do not depend on properties of the used ``thermometer'' such as tracer size and mass or a confining potential, and (iii) mutually lead to similar temperature values over a wide parameter regime. In particular, the kinetic temperatures $T_{\text{kin}}$, $T_{\text{kin4}}$ and the configurational temperature $T_{\text{conf}}$ have these advantages.

The present study serves as a starting point towards a systematic classification and unification of different possibilities to define temperature in active systems. It invites further studies to generalize the suggested temperature definitions and to fundamentally explain and exploit the identified temperature classes that lead to similar temperature values. Such studies could also answer the fundamental question of which of the presented temperatures can be interpreted as a measure for the direction of energy transfer as heat. Of course, alternatively, for non-homogeneous systems, one can choose to give up the definition of a global temperature altogether in far-from-equilibrium systems and to define a local temperature instead, which can be done by calculating the presented temperature definitions in a subdomain of the considered system.

%%%%%%%%%%%%%%%%%%%%%%%%%%%%%%%%%%%%%%%%%%%%%%%%%%%%%%%%%%%%%%%%%%%%%%%%%%%%%%%%%%%%%%%%%%%%%%%%%%%%%%%%%%%%%
%%%%%%%%%%%%%%%%%%%%%%%%%%%%%%%%%%%%%%%%%%%%%%%%%%%%%%%%%%%%%%%%%%%%%%%%%%%%%%%%%%%%%%%%%%%%%%%%%%%%%%%%%%%%%
% BACKMATTER
%%%%%%%%%%%%%%%%%%%%%%%%%%%%%%%%%%%%%%%%%%%%%%%%%%%%%%%%%%%%%%%%%%%%%%%%%%%%%%%%%%%%%%%%%%%%%%%%%%%%%%%%%%%%%
%%%%%%%%%%%%%%%%%%%%%%%%%%%%%%%%%%%%%%%%%%%%%%%%%%%%%%%%%%%%%%%%%%%%%%%%%%%%%%%%%%%%%%%%%%%%%%%%%%%%%%%%%%%%%
% \backmatter
\section*{Acknowledgments}
The authors thank Jeppe Dyre for useful discussions. L.H.\ gratefully acknowledges the support by the German Academic Scholarship Foundation (Studienstiftung des deutschen Volkes). H.L.\ was supported by the German Research Foundation (DFG) within project LO 418/29-1.

\section*{Author declarations}
\subsection*{Conflict of interest statement}
The authors have no conflicts to disclose.

\subsection*{Author contributions}
\textbf{Lukas Hecht:} data curation, formal analysis, investigation, software, visualization, writing (original draft and review \& editing). \textbf{Lorenzo Caprini:} investigation, writing (original draft and review \& editing). \textbf{Hartmut L\"owen:} conceptualization, writing (review \& editing). \textbf{Benno Liebchen:} conceptualization, supervision, writing (original draft and review \& editing).

\section*{Data availability}
The data that support the findings of this study are available from the corresponding author upon reasonable request.

%%%%%%%%%%%%%%%%%%%%%%%%%%%%%%%%%%%%%%%%%%%%%%%%%%%%%%%%%%%%%%%%%%%%%%%%%%%%%%%%%%%%%%%%%%%%%%%%%%%%%%%%%%%%%
%%%%%%%%%%%%%%%%%%%%%%%%%%%%%%%%%%%%%%%%%%%%%%%%%%%%%%%%%%%%%%%%%%%%%%%%%%%%%%%%%%%%%%%%%%%%%%%%%%%%%%%%%%%%%
% APPENDICES
%%%%%%%%%%%%%%%%%%%%%%%%%%%%%%%%%%%%%%%%%%%%%%%%%%%%%%%%%%%%%%%%%%%%%%%%%%%%%%%%%%%%%%%%%%%%%%%%%%%%%%%%%%%%%
%%%%%%%%%%%%%%%%%%%%%%%%%%%%%%%%%%%%%%%%%%%%%%%%%%%%%%%%%%%%%%%%%%%%%%%%%%%%%%%%%%%%%%%%%%%%%%%%%%%%%%%%%%%%%
\appendix
\renewcommand{\theequation}{A\arabic{equation}}		
\setcounter{equation}{0}

%%%%%%%%%%%%%%%%%%%%%%%%%%%%%%%%%%%%%%%%%%%%%%%%%%%%%%%%%%%%%%%%%%%%%%%%%%%%%%%%%%%%%%%%%%%%%%%%%%%%%%%%%%%%%
\section*{Appendix A: The active Ornstein-Uhlenbeck particle model}
The active Ornstein-Uhlenbeck particle (AOUP) model represents an alternative model to the ABP model and is commonly used for analytical calculations.\cite{Nguyen_JPhysCondensMatter_2022,Hecht_ArXiv_2021,Bonilla_PhysRevE_2019,Martin_PhysRevE_2021} Similar to the ABP model, the AOUP model is a dry model and the translational degrees of freedom follow the Langevin equation given in Eq.\ (1) in the main text. In contrast, in the case of AOUPs moving in two spatial dimensions, the orientation vector $\mathbf{p}_i$ is represented by a two-dimensional Ornstein-Uhlenbeck process that allows both the modulus $p_i$ and the orientation angle $\phi_i$ to fluctuate with related amplitudes. The dynamics of $\mathbf{p}_i$ is described by
\begin{equation}
	\frac{\text{d}\mathbf{p}_i}{\text{d}t}= - \frac{\mathbf{p}_i}{\tau_{\text{p}}} + \sqrt{\frac{1}{\tau_{\text{p}}}}\boldsymbol{\chi}_i, \label{eq:aoup-rot-od}
\end{equation}
where again $\tau_{\text{p}}=1/D_{\text{r}}$ denotes the persistence time and $\boldsymbol{\chi}_i$ is Gaussian white noise with zero mean and unit variance.\cite{Caprini_JCP_2021,Hecht_ArXiv_2021,Nguyen_JPhysCondensMatter_2022} Note that with the notation used here, both the ABP and the AOUP model share the same autocorrelation function of the orientation vector $\mathbf{p}_i$ with an exponential shape, i.e., $\langle\mathbf{p}_i(t)\cdot\mathbf{p}_i(0)\rangle = \exp(-t/\tau_{\text{p}})$, and thus, also the same equal-time second moment $\langle\mathbf{p}_i^2\rangle=1$.\cite{Farage_PhysRevE_2015,Caprini_JCP_2022} The difference between the two models is visible in the higher-order moments and the full-shape of the active-force distribution. Indeed, the latter is Gaussian in the case of AOUPs but it is characterized by a constant modulus in case of ABPs.\cite{Caprini_JCP_2022}

%%%%%%%%%%%%%%%%%%%%%%%%%%%%%%%%%%%%%%%%%%%%%%%%%%%%%%%%%%%%%%%%%%%%%%%%%%%%%%%%%%%%%%%%%%%%%%%%%%%%%%%%%%%%%
\section*{Appendix B: Analytical results for the kinetic temperature}
In the main text, we discussed the analytical expression of the kinetic temperature for free non-interacting ABPs. Let us now consider a few more complicated setups. By confining the system through an external harmonic potential $U_{\text{ext}}(\mathbf{r})=k\mathbf{r}^2/2$, for the AOUP case [Eq.\ (\ref{eq:aoup-rot-od})], one obtains \cite{Caprini_JCP_2021}
\begin{equation}
	\alpha = \frac{\tau_{\text{p}}\gamma_{\text{t}}/m}{1+\tau_{\text{p}}\gamma_{\text{t}}/m + \tau_{\text{p}}^2 k / m}. \label{eq:Tkin-alpha-harmpot}
\end{equation}
Since the second moments of the distribution for ABPs and AOUPs are equal, this result holds also for ABPs. Equation (\ref{eq:Tkin-alpha-harmpot}) shows that the harmonic confinement reduces the kinetic temperature. In the overdamped regime, i.e., $\tau_{\text{p}}\gamma_{\text{t}}/m\gg 1$, Eq.\ (\ref{eq:Tkin-alpha-harmpot}) simplifies to
\begin{equation}
	\alpha = \frac{1}{1+\tau_{\text{p}}k/\gamma_{\text{t}}}.\label{eq:Tkin-alpha-harmpot-od}
\end{equation}
For a general external potential $U_{\text{ext}}$, exact analytical results are not known. However, naively, we can derive an approximate result based on an equilibrium-like approximation obtained in the overdamped regime, which reads \cite{Caprini_SciRep_2019}
\begin{equation}
	\alpha \approx \left[1 + \frac{\tau_{\text{p}}}{\gamma_{\text{t}}} \nabla^2 U_{\text{ext}}(\mathbf{r}) \right]^{-1}.\label{eq:Tkin-alpha-generalpot}
\end{equation}
Note that this result is consistent with Eq.\ (\ref{eq:Tkin-alpha-harmpot-od}) for the harmonic external potential in the overdamped regime.

For interacting active particles, there are no simple analytical expressions for the kinetic temperature except in very dense systems displaying a solid configuration. In this case, for AOUPs, we obtain \cite{Caprini_SoftMatter_2021}
\begin{equation}
	\kb T_{\text{kin}}^{\text{AOUP}} = \kb T_{\text{b}} + \frac{v_0^2\tau_{\text{p}}\gamma_{\text{t}}}{1 + \tau_{\text{p}}/\tau_I + 6 \omega_E^2 \tau_{\text{p}}^2} \frac{\mathcal{I}}{\pi}.
\end{equation}
The term $\omega_{\text{E}}^2$ reads
\begin{equation}
	\omega_{\text{E}}^2 = \frac{1}{2m} \left( u''(\bar{x}) + \frac{u'(\bar{x})}{\bar{x}}  \right),
\end{equation}
where $\bar{x}$ is the average distance between different particles, i.e., the lattice constant of the crystal, $u$ is the interaction potential, and $\mathcal{I}$ is a numerical factor that shows a non-trivial dependence on $\tau_{\text{p}}$, $m/\gamma_{\text{t}}$, and $\omega_{\text{E}}$.\cite{Caprini_SoftMatter_2021} We remark that the kinetic energy of a single active particle in a solid configuration is smaller than the kinetic energy of a free active particle: In the solid, the neighboring particles hinder the motion of a target particle and decrease its kinetic energy.

%%%%%%%%%%%%%%%%%%%%%%%%%%%%%%%%%%%%%%%%%%%%%%%%%%%%%%%%%%%%%%%%%%%%%%%%%%%%%%%%%%%%%%%%%%%%%%%%%%%%%%%%%%%%%
\section*{Appendix C: Virial temperature for inertial ABPs}
For inertial ABPs [Eqs.\ (\ref{eq:active-trans-leq}) and (\ref{eq:abp-rot-leq})], the virial temperature can be calculated by inserting the total force, i.e., the right-hand side of Eq.\ (\ref{eq:active-trans-leq}). This leads to four contributions: The first contribution comes from the drag force $-\gamma_{\text{t}}\mathbf{v}_i$ and involves the correlation function $\langle\mathbf{r}_i\cdot\mathbf{v}_i\rangle$. In the steady state, it can be rewritten in terms of the effective diffusion coefficient:
\begin{equation}
	2\langle\mathbf{r}_i\cdot\mathbf{v}_i\rangle=\partial_t\langle\mathbf{r}_i\cdot\mathbf{r}_i\rangle=\partial_t\text{MSD}(t)\xrightarrow[t\gg 1]{} \partial_t (2dD_{\text{eff}}t)=2dD_{\text{eff}}.
\end{equation}
This contribution is equal to the Einstein temperature as defined in Eq.\ (\ref{eq:einstein-temperature}). The second contribution comes from the effective self-propulsion force $\gamma_{\text{t}}v_0\mathbf{p}_i$. It contains the correlation $\langle\mathbf{r}_i\cdot\mathbf{p}_i\rangle$ between the position and orientation of the active particles. The third contribution involves the Gaussian white noise $\sqrt{2\kb T_{\text{b}}\gamma_{\text{t}}}\boldsymbol{\xi}_i$. It leads to the correlation $\langle\mathbf{r}_i\cdot\boldsymbol{\xi}_i\rangle=0$.\cite{Sandoval_PhysRevE_2020} The remaining contribution involves the interaction forces $\mathbf{F}_{ij}=-\boldsymbol{\nabla}_{\mathbf{r}_i}u(r_{ij})$ and possible external forces. The contribution from the interaction forces can be written as
\begin{align}
	\sum_{i=1}^{N}\langle\mathbf{r}_i\cdot\mathbf{F}_i\rangle&=\sum_{i=1}^{N}\sum_{j=1,j\neq i}^{N}\langle\mathbf{r}_i\cdot\mathbf{F}_{ij}\rangle\nonumber\\
	&=\sum_{i=2}^{N}\sum_{j=1}^{i-1}\mathbf{r}_i\cdot\mathbf{F}_{ij} +	\sum_{i=2}^{N}\sum_{j=1}^{i-1}\mathbf{r}_j\cdot\mathbf{F}_{ji}\nonumber\\
	&=\sum_{i=2}^{N}\sum_{j=1}^{i-1}(\mathbf{r}_i-\mathbf{r}_j)\cdot\mathbf{F}_{ij}\nonumber\\
	&=\sum_{i=1}^{N}\sum_{j<i}\langle\mathbf{r}_{ij}\cdot\mathbf{F}_{ij}\rangle
\end{align}
by applying Newton's third law and using $\mathbf{F}_i=\sum_{j=1,j\neq i}^{N}\mathbf{F}_{ij}$. Here, we use $\mathbf{r}_{ij}=\mathbf{r}_i-\mathbf{r}_j$. Finally, we can write the virial temperature for inertial ABPs as
\begin{equation}
	\kb T_{\rm vir}^{\text{ABP}}=\gamma_{\text{t}}\lim\limits_{t\rightarrow\infty}\partial_t\text{MSD}(t) + \frac{1}{2Nd}\sum_{i=1}^{N} \left\langle\mathbf{r}_i\cdot\mathbf{F}_{\text{ext},i} - \sum_{j<i}\mathbf{r}_{ij}\cdot\mathbf{F}_{ij} - \gamma_{\text{t}}v_0\mathbf{r}_i\cdot\mathbf{p}_i \right\rangle.
\end{equation}

%%%%%%%%%%%%%%%%%%%%%%%%%%%%%%%%%%%%%%%%%%%%%%%%%%%%%%%%%%%%%%%%%%%%%%%%%%%%%%%%%%%%%%%%%%%%%%%%%%%%%%%%%%%%%
\section*{Appendix D: Derivation of the oscillator temperature}
Consider a passive tracer particle trapped in a harmonic potential and suspended in a bath of Brownian particles. Due to the collisions of the bath particles with the tracer, the latter is driven by these collisions, which can be modeled as random driving force following a Gaussian white noise process. Let $x$ denote the displacement of the tracer particle with respect to its equilibrium position (here only in one spatial dimension for simplicity). Then, the equation of motion for the tracer particle of mass $m$ reads
\begin{equation}
	m\ddot{x}=-\gamma\dot{x} - kx + \sqrt{2\kb T\gamma}\xi(t)\label{eq:harm-osc-noise-driven}
\end{equation}
with the drag coefficient $\gamma$ of the bath, the force constant $k$, and Gaussian white noise $\xi(t)$ of zero mean and unit variance. Here, $T$ denotes the temperature of the bath, which is related to the position fluctuations $\erw{x^2}$ via
\begin{equation}
	\erw{x^2}=\frac{\kb T}{k}.
\end{equation}
This relation can be derived as follows: First, we write down the Fokker-Planck equation for the probability density $\mathcal{P}(x,v,t)$ with $v=\dot{x}$ by following the standard text book:\cite{Risken_Book_TheFokker-PlanckEquation_1984}
\begin{equation}
	\pder{\mathcal{P}}{t}=\left\lbrace\frac{\gamma}{m}-v\pder{}{x}+\left(\frac{\gamma}{m}v+\frac{k}{m}x\right)\pder{}{v}+\frac{\kb T\gamma}{m^2}\frac{\partial^2}{\partial v^2} \right\rbrace\mathcal{P}.
\end{equation}
Now, it can be shown that the solution of the Fokker-Planck equation is given by \footnote{The solution can be derived from the condition $\partial_t\mathcal{P}=0$ by using a Gaussian distribution as an ansatz for $\mathcal{P}(x,v)$. It is also given in Ref.\ \onlinecite{Dybiec_PhysRevE_2017}.} 
\begin{equation}
	\mathcal{P}(x,v)=\frac{1}{A}\ex{-\frac{1}{\kb T}\left(\frac{1}{2}mv^2+\frac{1}{2}kx^2\right)},
\end{equation}
which is simply the Boltzmann distribution with normalization constant $A=\Int{}{}{x}\Int{}{}{v}\mathcal{P}(x,v)$.\cite{Boltzmann_WienBer_1868} Then, the position fluctuations can be determined as
\begin{align}
	\erw{x^2}&=\frac{\Int{}{}{x}\Int{}{}{v}x^2\mathcal{P}(x,v)}{\Int{}{}{x}\Int{}{}{v}\mathcal{P}(x,v)}\nonumber\\
	&=\frac{\Int{}{}{x}x^2\ex{-\frac{kx^2}{2\kb T}}}{\Int{}{}{x}\ex{-\frac{kx^2}{2\kb T}}}\nonumber\\
	&=\frac{\sqrt{\pi}}{2}\left(\frac{2\kb T}{k}\right)^{(3/2)}\sqrt{\frac{k}{2\pi\kb T}}=\frac{\kb T}{k},
\end{align}
where we have used Eqs.\ (21.24b) and (21.25) from Ref.\ \onlinecite[p.\ 1100]{Bronshtein_Book_HandbookOfMathematics_2015}.

%%%%%%%%%%%%%%%%%%%%%%%%%%%%%%%%%%%%%%%%%%%%%%%%%%%%%%%%%%%%%%%%%%%%%%%%%%%%%%%%%%%%%%%%%%%%%%%%%%%%%%%%%%%%%
\section*{Appendix E: The effective temperature}
The effective temperature is based on linear response theory and the fluctuation dissipation theorem (FDT). It can be derived as follows: Let us consider a (weak) time-dependent perturbation that couples to an observable $A$. Then, the linear response function, which describes the response of an observable $B$ to the time-dependent perturbation, is given by 
\begin{equation}
	R_{AB}(t,t')=-\frac{1}{\kb T}\erw{\dot{B}(t)A(t')},
\end{equation}
where the average $\erw{\cdot}$ is taken over the unperturbed system.\cite{Hansen_Book_TheoryOfSimpleLiquids_2006} It is related to the time-integrated linear response (susceptibility) by
\begin{equation}
	\chi_{AB}(t,0)=\Int{0}{t}{t'}R_{AB}(t,t'),
\end{equation}
following the notation in Ref.\ \onlinecite{Cugliandolo_FluctNoiseLett_2019}. By setting $A=B=x$, where $x$ denotes the position of a particle in $x$ direction, one can show that
\begin{align}
	\chi_{xx}(t,0)&=\Int{0}{t}{t'}R_{xx}(t,t')\nonumber\\
	&=\frac{1}{\kb T}\Int{0}{t}{t'}\Int{0}{t'}{t''}\langle\dot{x}(t')\dot{x}(t'')\rangle\nonumber\\
	&=\frac{1}{2\kb T}\Int{0}{t}{t'}\Int{0}{t}{t''}\langle\dot{x}(t')\dot{x}(t'')\rangle\nonumber\\
	&=\frac{1}{2\kb T}{\rm MSD}(t)
\end{align}
with the mean-square displacement ${\rm MSD}(t)=\langle\left[x(t)-x(0)\right]^2\rangle$. Hence, the FDT for the time-integrated linear response reads
\begin{equation}
	2\kb T\chi_{xx}(t)={\rm MSD}(t).
\end{equation}
Following Ref.\ \onlinecite{Cugliandolo_FluctNoiseLett_2019}, this can be generalized to $d$ spatial dimensions:
\begin{equation}
	2d\kb T\chi(t)={\rm MSD}(t),
\end{equation}
with
\begin{equation}
	\chi(t)=\Int{0}{t}{t''}\Sum{\alpha=1}{d}R_{\alpha\alpha}(t,t'').
\end{equation}
In order to define an effective temperature for systems out of equilibrium, one introduces a time-dependent effective temperature $T_{\rm eff}(t)$, which is defined by \cite{Cugliandolo_FluctNoiseLett_2019,Petrelli_PhysRevE_2020}
\begin{equation}
	\kb T_{\rm eff}(t)=\frac{\text{MSD}(t)}{2d\chi(t)}.
\end{equation}

%%%%%%%%%%%%%%%%%%%%%%%%%%%%%%%%%%%%%%%%%%%%%%%%%%%%%%%%%%%%%%%%%%%%%%%%%%%%%%%%%%%%%%%%%%%%%%%%%%%%%%%%%%%%%
%%%%%%%%%%%%%%%%%%%%%%%%%%%%%%%%%%%%%%%%%%%%%%%%%%%%%%%%%%%%%%%%%%%%%%%%%%%%%%%%%%%%%%%%%%%%%%%%%%%%%%%%%%%%%
% REFERENCES
%%%%%%%%%%%%%%%%%%%%%%%%%%%%%%%%%%%%%%%%%%%%%%%%%%%%%%%%%%%%%%%%%%%%%%%%%%%%%%%%%%%%%%%%%%%%%%%%%%%%%%%%%%%%%
%%%%%%%%%%%%%%%%%%%%%%%%%%%%%%%%%%%%%%%%%%%%%%%%%%%%%%%%%%%%%%%%%%%%%%%%%%%%%%%%%%%%%%%%%%%%%%%%%%%%%%%%%%%%%
\clearpage
%aipnum4-2.bst 2019-01-14 (MD) hand-edited version of apsrev4-1.bst

\end{document}